\documentclass[twocolumn]{aastex62}

\usepackage{amsmath}

\usepackage{bm}
\usepackage{newtxtext,newtxmath}
\usepackage{color}
\usepackage{tensor}
\usepackage{multirow}
\usepackage{graphicx}
\usepackage{natbib}
\usepackage{enumerate}
\usepackage[english]{babel}
\usepackage{tabularx}
\usepackage{placeins}
\usepackage{url}
\usepackage[caption=false]{subfig}
\usepackage{layouts}
\usepackage{ulem}
\usepackage{comment}
\usepackage{needspace}

\newcommand{\Msol}{\mathrm{M}_{\odot}}

\newcommand{\Mpc}{\mathrm{Mpc}}

\def\lsim{\mathrel{\rlap{\lower3pt\hbox{$\sim$}}
		\raise1pt\hbox{$<$}}}                
\def\gsim{\mathrel{\rlap{\lower3pt\hbox{$\sim$}}
		\raise1pt\hbox{$>$}}}                

\renewcommand{\vec}[1]{\bm{#1}}
\newcommand{\vdelta}{\vec{\delta}}

\newcommand{\codefont}[1]{{\texttt{#1}}}
\newcommand{\genetIC}{\codefont{genetIC}} 
\newcommand{\svs}[1]{{\color{orange}}}

\newcommand{\classname}[1]{\texttt{#1}}

\newcommand{\matrixfont}[1]{{\mathsf{#1}}}
\newcommand{\threevec}[1]{\bm{\mathit{#1}}}
\newcommand{\nvec}[1]{\bm{\mathit{#1}}}

\begin{document}
	
\title{\codefont{GenetIC} --- a new initial conditions generator to support genetically modified zoom simulations}

\date{ Received ---; published---. }
\correspondingauthor{Stephen Stopyra}
\author{Stephen Stopyra}
\affiliation{Department of Physics and Astronomy, University College London, London WC1E 6BT, UK}
\email{s.stopyra@ucl.ac.uk}
\author{Andrew Pontzen}
\affiliation{Department of Physics and Astronomy, University College London, London WC1E 6BT, UK}
\author{Hiranya Peiris}
\affiliation{Department of Physics and Astronomy, University College London, London WC1E 6BT, UK}
\affiliation{The Oskar Klein Centre for Cosmoparticle Physics, Department of Physics, Stockholm University, AlbaNova, Stockholm SE-106 91, Sweden}
\author{Nina Roth}
\affiliation{Department of Physics and Astronomy, University College London, London WC1E 6BT, UK}
\affiliation{Zurich Gruppe Deutschland, Deutzer Allee 1, 50679 K\"{o}ln, Germany}
\author{Martin P. Rey}
\affiliation{Lund Observatory, Department of Astronomy and Theoretical Physics, Lund University, Box 43, SE-221 00, Lund, Sweden}
\affiliation{Department of Physics and Astronomy, University College London, London WC1E 6BT, UK}

\begin{abstract}
	
	We present \codefont{genetIC}, a new code for generating initial conditions for cosmological $N$-body simulations. The code allows precise, user-specified alterations to be made to arbitrary regions of the simulation (while maintaining consistency with the statistical ensemble). These ``genetic modifications'' allow, for example, the history, mass, or environment of a target halo to be altered in order to study the effect on their evolution. The code natively supports initial conditions with nested zoom regions at progressively increasing resolution. Modifications in the high-resolution region must propagate self-consistently onto the lower resolution grids; to enable this while maintaining a small memory footprint, we introduce a Fourier-space filtering approach to generating fields at variable resolution. Due to a close correspondence with modifications, constrained initial conditions can also be produced by \codefont{genetIC} (for example with the aim of matching structures in the local Universe).  We test the accuracy of modifications performed within zoom initial conditions. The code achieves sub-percent precision, which is easily sufficient for current applications in galaxy formation.
	\vspace{1cm}	
\end{abstract}

\section{Introduction} \label{sec:intro}

The generation of initial conditions is a crucial step in simulations of cosmological structure and galaxy formation. Simulation codes require as input the positions and velocities  of dark matter and baryons at an early time in the Universe's development, when deviations from homogeneity are approximately linear. The basic task is to generate a Gaussian random field, with a specific power spectrum, on a discrete grid that samples the continuous density-contrast field. This in turn can be used to generate velocity and displacement fields for a set of particles and, for mesh codes, fluid variables for each grid cell. Starting from Gaussian white noise, all these fields can be generated via a suitable series of convolutions.

A significant complication in performing these convolutions arises if we wish to work with zoomed initial conditions, where the grid spacing differs from one region of the simulation to another. Zoom simulations are attractive because they focus computational effort on a single object, allowing it to be modeled with far greater fidelity than is possible for a population; but, for initial condition generation, efficient convolution algorithms typically require a fixed grid resolution.  One possibility is to generate the field at uniformly high resolution, and down-sample within the unzoomed majority of the domain \citep[e.g.][]{katz1994formation,navarro1994simulations,tormen1997structure,2008ApJS..178..179P}. However this becomes prohibitively wasteful of memory and processing resources as the desired dynamic range between zoom and volume increases. As a solution to this problem,  \codefont{GRAFIC2} \citep{2001ApJS..137....1B} introduced an algorithm that generates fields directly on a nested grid structure, albeit with the need to pad out the high resolution region to twice its final side length. A refined implementation of the same basic algorithm is provided by the widely-used generator \codefont{MUSIC} \citep{hahn2011multi}. 

These initial conditions generators have helped zooms become a standard technique for high resolution galaxy formation studies. However, we often want to understand how a galaxy's observable properties have been affected by its history and local environment. This is impossible with a single object or even a small sample from a typical cosmological volume, given that each galaxy differs from its counterparts in multiple distinct respects. A ``brute force'' approach of generating many different random realizations of the field until we find one that looks sufficiently similar -- yet with the desired differences -- would be exceptionally time-consuming and imprecise.  Consequently, an attractive alternative is to generate systematic variations in the accretion history, environment, or other aspect of a {\it single} object which one then resimulates several times. In this paper we will describe a code (available at \url{https://github.com/pynbody/genetIC}) which generates and then minimally modifies initial conditions to make such ``genetic modifications'', while maximizing the likelihood for the field to have arisen as a random Gaussian draw from the $\Lambda$CDM power spectrum \citep{Roth:2015wha,Pontzen:2016wwf,2019MNRAS.485.1906R}.

Operationally, such modifications closely resemble the Hoffman-Ribak procedure for generating constrained initial conditions \citep{hoffman1991constrained}. Existing initial conditions codes which have the ability to perform Hoffman-Ribak constraints can therefore in principle be used for simple modifications as well. However, in existing implementations the algorithm can only be applied to a single grid at once; when applied to a zoom simulation, modifications do not propagate correctly out of the high resolution region, leading to discontinuities and errors in the correlation function. Moreover, an effective modification algorithm needs to target the field value averaged in regions of arbitrary shape, as determined by tracing the material in halos or their environment back to the initial conditions \citep{Roth:2015wha}. To our knowledge no public algorithm for performing such multi-resolution or arbitrary shape manipulations currently exists.

In this paper we present \codefont{genetIC}, a new code which implements solutions to these issues and so is suitable for generating zoomed, genetically modified initial conditions. As a fortuitous side-effect of implementing multi-resolution modifications, \codefont{genetIC} also drastically reduces the memory footprint for realizations on a given zoom geometry, due to near-elimination of the need for padding.  At present, output can be made to \codefont{GADGET}, \codefont{TIPSY} or \codefont{GRAFIC} formats (the latter being suitable for use with \codefont{RAMSES}). While not its primary focus, \codefont{genetIC} can also perform global field manipulations such as inversion and power spectrum fixing, which enable insights into the growth of large scale structure \citep[e.g.][]{PhysRevD.93.103519,Angulo:2016hjd,Anderson19}. The code is modular and extensible so that additional manipulations or output formats can easily be added at a later date.

We review the generation of multi-resolution initial conditions and explain the new implementation in \codefont{genetIC} in Section \ref{sec:desc}, with supporting mathematical derivations given in Appendix~\ref{app:technical}. The core algorithm used for applying constraints consistently across resolution boundaries is described in Section \ref{sec:constraints}, with the full details  in Appendix \ref{app:constraints}. We discuss the accuracy of \codefont{genetIC} with examples in Section \ref{sec:examples}. Section \ref{sec:overview} then gives an overview of the structure of the code. A short summary is provided in Section \ref{sec:disc}.

\needspace{6em}
\section{Description of genetIC}
\label{sec:desc}

\subsection{Review of Generating Initial Conditions\label{sec:ics_review}}

At its heart, the problem that any initial conditions generator aims to solve is generating a set of $N$ particles that approximate a smooth density field described by an initial density contrast, $\delta(\threevec{x})$. Genetic modification, which we will describe in Section \ref{sec:constraints}, seeks to manipulate this field in a manner consistent with a draw from the same Gaussian ensemble, as specified by the cosmological power spectrum.

The typical procedure is first to generate a discrete vector, $\nvec{\delta}$, that samples $\delta(\nvec{x})$ at a set of points $\vec{x}_i$ on a regular lattice. The  vector $\nvec{\delta}$ should be a random draw from a distribution with a covariance determined by the cosmology, that is
\begin{equation}
\langle\nvec{\delta} \nvec{\delta}^{\dagger}\rangle = \matrixfont{C}\label{eq:cov},
\end{equation}
where $\matrixfont{C}$ is a discrete version of the cosmological covariance matrix. To create $\nvec{\delta}$ we start from a unit-variance, uncorrelated white-noise field $\nvec{n}$, and then multiply by $\matrixfont{C}^{1/2}$.

Once generated, each element of the vector $\nvec{\delta}$ gives the density contrast averaged over a particular grid cell of the simulation. The next task is to translate this into a corresponding set of particles with positions and velocities. This can be achieved using Lagrangian perturbation theory \citep[see][]{1993A&A...267L..51B,1993MNRAS.264..375B}. In this prescription, particles are labeled by their initial grid positions, $\nvec{q}$, and their evolution is tracked using a displacement field, $\nvec{\Psi}(\tau,\nvec{q})$ that gives their position at some later time; the dynamics of $\nvec{\Psi}(\tau,\nvec{q})$ can be solved perturbatively. The lowest order terms constitute the \cite{Zeldovich:1969sb} approximation, which links velocities and displacements directly to gradients of the potential. Provided that we start at sufficiently high redshift (typically $z >100$), the Zel'dovich approximation is adequate for galaxy formation questions for which \codefont{genetIC} is principally designed; however the modular nature of \codefont{genetIC} allows the method used to be extended to higher order if required --- see Section \ref{sec:overview} for further discussion.\\

\needspace{3em}
\subsection{Zoom Initial Conditions\label{sec:zoom_review}}

\begin{figure}
	\centering
	\includegraphics[width=0.45\textwidth]{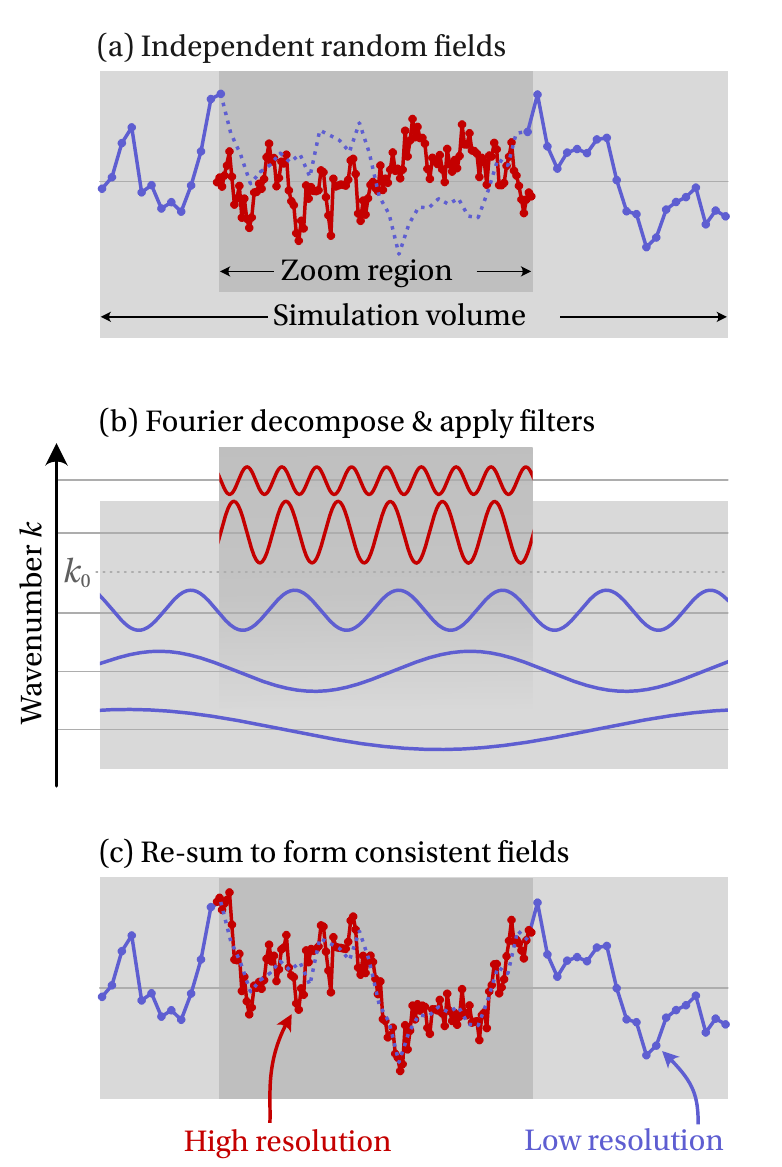}
	\caption{\label{fig:zoomset up} Illustration in 1D of zoom simulation initial conditions, as implemented by \codefont{genetIC}. A high-resolution grid is inserted as a ``zoom region'' into a low-resolution grid. (a)~Both grids are initially seeded independently, but this generates long-wavelength modes in the high-resolution grid that are inconsistent with the low-resolution field. To solve this problem, we (b) split modes in Fourier space, as described in Section~\ref{sec:fast_filter}. Only high-frequency modes (wavenumber above $k_0$) are retained on the new grid. (c) We then replace the missing long-wavelength modes with an appropriately filtered version of those in the low-resolution grid. }
\end{figure}

The prescription described in Section \ref{sec:ics_review} is complete for a uniform grid. However, most applications of genetic modifications will make use of zoom simulations incorporating the high resolution needed to simulate individual galaxies and halos, while accurately retaining the gravitational effects of a large scale environment \citep[e.g.][]{navarro1994simulations,katz1994formation}. This setup is illustrated in the bottom panel of Figure~\ref{fig:zoomset up}. The upper two panels show the field generation process in \codefont{genetIC}, which we will now motivate before describing in detail in Section~\ref{sec:fast_filter}. In the depicted example there are two grids; a low-resolution grid covering the full simulation domain, and a high-resolution grid which covers the ``zoom'' region. Any approach to generating zoom initial conditions must solve the problem of how to relate random fields on the two grids in order to obtain the correct correlation structure between the two regions.

One possible way to generate a zoom simulation is to calculate the entire cosmological volume at high resolution, then degrade the sampling outside the target region. However, this is inefficient, or for some applications entirely infeasible; for example in the \codefont{EDGE} suite \citep{2019ApJ...886L...3R,2020MNRAS.491.1656A} the initial conditions resolution corresponds to an effective grid of $32768^3$. Storing a single such grid at double precision requires 32 terabytes; manipulating it is entirely out of reach.
To obtain the high effective resolution of \codefont{EDGE}, a $1024^3$ grid is instead nested inside two $512^3$ grids. Each level has a physical extent four times smaller than the one above. A single field in this hierarchy requires only ten gigabytes to store, despite reaching the required effective resolution in the region of interest. 

As stated above, the challenge is to ensure the final, multi-resolution field has the correct correlations despite being represented in a piecemeal manner. Specifically, we need the long-wavelength modes of the fine grid to match those of the low-resolution grid in the region where they overlap, and for the finite box-size of the fine grid and the boundary between the two grids to have minimal impact.

The only solution to this problem currently described in the literature is presented by \citet{2001ApJS..137....1B} and \citet{hahn2011multi}. It involves nesting a large buffer region around the section of the high resolution grid before applying convolutions, in order to create the correct boundary conditions for the small grid,  so allowing long-wavelength convolutions to be carried out safely.  These ``ghost'' regions expand the total storage requirements to 73 gigabytes per field (in the \codefont{EDGE} example given above). Note that the components of the output displacement field, $\nvec{\Psi}(\tau,\nvec{q})$, each count as a field, making the minimal memory requirements of such an algorithm almost 300 gigabytes for this scenario. While sufficiently powerful computer resources are available, once modifications are introduced (each one with their own associated field), the computational demands spiral further upwards.  Moreover there is no existing algorithm describing how ghost regions should correctly interact with modifications to the field. This poses a problem since, to maintain continuity and consistency with the cosmological power spectrum it is essential that the long-wavelength behavior of the modification propagates correctly outside the zoom region.

The two considerations above motivate finding a solution to multi-resolution convolution and field modification that does not involve ghost regions. Our approach is to view the fundamental problem as one of band-limiting: we wish to supplant the original low-resolution information with additional modes above the original Nyquist frequency\footnote{The Nyquist frequency $k_{\mathrm{nyq}}$ is defined to be half the sampling rate, i.e. $k_{\mathrm{nyq}} = N\pi/L$ where $N$ is the number of grid points along one edge of the simulation box.}. From this perspective, the solution is to combine modes in Fourier space. Figure \ref{fig:zoomset up} outlines how this procedure starts from two independent random fields on the separate grids and combines them into a consistent multi-resolution realization. The Fourier perspective also leads to an algorithm for applying field modifications across resolution boundaries that will be discussed in more detail in Section \ref{sec:constraints}.

The \codefont{MUSIC} code \citep{hahn2011multi} has an option to use Fourier-space filtering, blending modes between low and high resolution grids during convolution with the correlation function (Hahn, private communication). This has not been explicitly documented in the scientific literature, but was first implemented for the \codefont{AGORA} code comparison project \citep{Agora16} and is now switched on by default. The \codefont{MUSIC} scheme is not designed to reduce memory requirements and so, compared to our approach, has a different set of design considerations for its Fourier filters. We will describe below how \codefont{genetIC} filters are band-limited in Fourier space but, unlike in \codefont{MUSIC}, also remain compact in real space.

\needspace{2cm}
\subsection{The Fourier-space Approach}
\label{sec:fast_filter}

\begin{figure*}
	\centering
	\includegraphics[width=\textwidth]{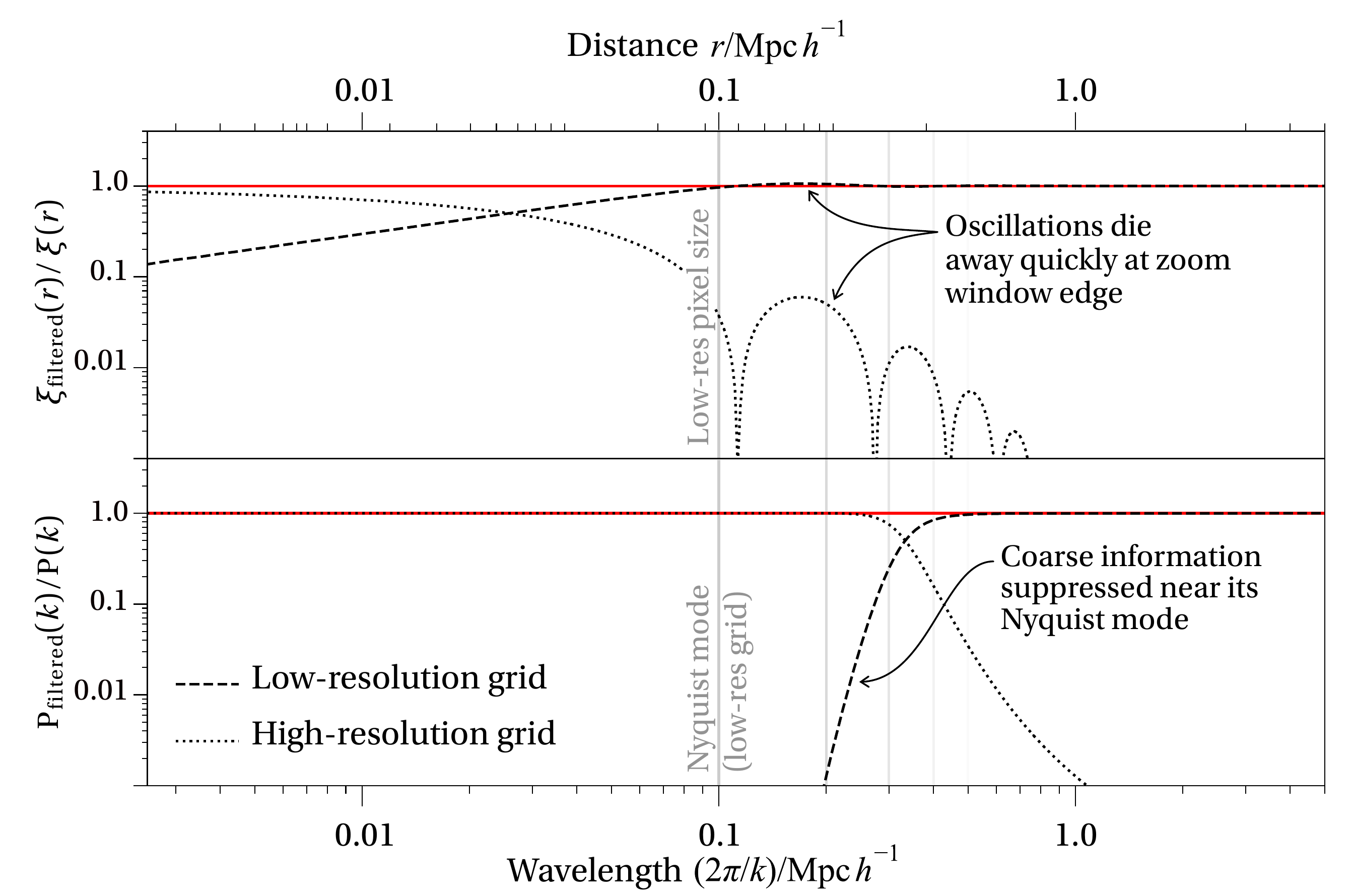}
	\caption{\label{fig:fermi-filter}Top: ratio of the filtered to unfiltered real-space correlation function, $\xi_{\mathrm{filtered}}(r)$, using the Fermi filter, equation~({\ref{eq:fermi}}), in Fourier space. The axes are logarithmic, so we show the absolute value of the ratio. The contribution from the high-resolution correlations dies away quickly (within a few low-resolution pixels), meaning the low-resolution field carries most of the correlation structure. However, over the decaying region there are some oscillations  due to the trade-off required between the sharpness of the filter in Fourier space and real-space (see Section \ref{sec:fast_filter} for further discussion).  Bottom: ratio of the filtered to unfiltered power spectrum for the same scenario. The wavelength $2\pi/k$ is used so that scales always increase from left to right. The filter is chosen to ensure we exclude small-scale modes above the Nyquist frequency of the low-resolution grid, while simultaneously ensuring the fine-grid, real-space correlation function rapidly decays outside the high-resolution region. Vertical grey lines show the low-resolution pixel size (top panel) and, equivalently, the Nyquist frequency (bottom panel) and its integer multiples.}
\end{figure*}
We will now describe in more detail how the procedure motivated above works in practice. We continue to focus on the case of two levels; nesting multiple subgrids requires recursively applying this two-level case. 

The initial aim is to construct $\nvec{\delta}_L$, a vector containing a low-resolution sampling of the density field on a coarsely-pixelized grid, and $\nvec{\delta}_H$, a high-resolution vector which stores information only in the zoom region. The challenge is to generate these starting from two independent white noise fields, $\nvec{n}_L$ and $\nvec{n}_H$ on the low-resolution and high-resolution grids respectively\footnote{The current version of \genetIC\ uses  independent white noise on each grid. An alternative would be to start from realizations which are already strongly correlated, so that the phase of high-frequency modes is not dependent on the placement of the zoom region.  This can be arranged, for example, by producing white noise relative to an octree basis function \citep{Jenkins:2013raa,Jenkins:2013nza}.}. 

In our Fourier-space approach, we think of this problem as being closely related to constructing filtered versions of an underlying high resolution field $\nvec{\delta}$ for the whole grid to which we do not have access. Such filtered fields would take the form
\begin{align}
\nvec{\tilde{\delta}}_L =&\, \matrixfont{F}_L\nvec{\delta},\label{eq:FL}\\
\nvec{\tilde{\delta}}_H =&\, \matrixfont{F}_H\nvec{\delta},\label{eq:FH}
\end{align}
where $\matrixfont{F_H}$ and $\matrixfont{F_L}$ are filters preserving high-frequency and low-frequency Fourier modes, respectively. By itself, this does not lead to a practical algorithm since $\nvec{\tilde{\delta}}_{L}$ and $\nvec{\tilde{\delta}}_H$ consume the same space in memory as the original vector, $\nvec{\delta}$. Instead we want $\nvec{\delta}_{L}$ to be a  pixelized version of $\nvec{\tilde{\delta}}_L$, defined on the low-resolution grid, and $\nvec{\delta}_H$ to be a version of $\nvec{\tilde{\delta}}_H$, which is  confined to the high-resolution region only. By choosing $\matrixfont{F}_L$ appropriately, $\nvec{\tilde{\delta}}_L$ is band-limited even on the coarse pixelization and therefore can be losslessly ``compressed'' to low resolution. Storing $\nvec{\delta}_H$ only in the zoom region can also be regarded as a compression, albeit one that explicitly destroys information which we do not require. 

This Fourier-space decomposition suggests a practical three-step algorithm for generating $\nvec{\delta}_L$ and $\nvec{\delta}_H$:
\begin{enumerate}[(a)]
\item Draw independent white noise fields for $\nvec{n}_L$ and $\nvec{n}_H$, and convolve with the theory cosmological correlation function independently on both grids to produce $\nvec{\delta}_L$ and $\nvec{\delta}_H$;
\item Apply the high-pass filter to $\nvec{\delta}_H$ and the low-pass filter to $\nvec{\delta}_L$; 
\item Sum the two fields to produce the final field in the high-resolution region. 
\end{enumerate}
This process is illustrated in Figure~\ref{fig:zoomset up}. We will discuss precisely how the fields from the two grids are combined in Appendix \ref{app:technical}. Note that, in step (a), convolving with the correlation function entails a trade-off between real-space and Fourier space accuracy \citep{Pen:1997up,Sirko:2005uz}. For uniform-resolution volumes this is a matter of preference \citep{orban2013keeping}, but zoom regions generated with the \cite{2001ApJS..137....1B} algorithm are only self-consistent when using a real-space transfer function \citep{hahn2011multi}. The \codefont{genetIC} algorithm computes all long-wavelength correlations on the base grid, which permits self-consistent convolution with either a real-space or Fourier-space transfer function, as desired.

We now turn to the specific choice of filter. The filters $\matrixfont{F}_L$ and $\matrixfont{F}_H$ need to be designed such that $\nvec{\delta}_L$ can be stored at the low pixel resolution without aliasing, while $\nvec{\delta}_H$ can be stored in the high resolution region and contains no large-scale correlations (so that the finite size of the high resolution region does not impinge on any manipulations). These requirements are in tension but can be satisfied approximately with an appropriate Fourier space filter. 

\begin{figure*}
	\begin{center}
		\includegraphics[width=0.8\textwidth]{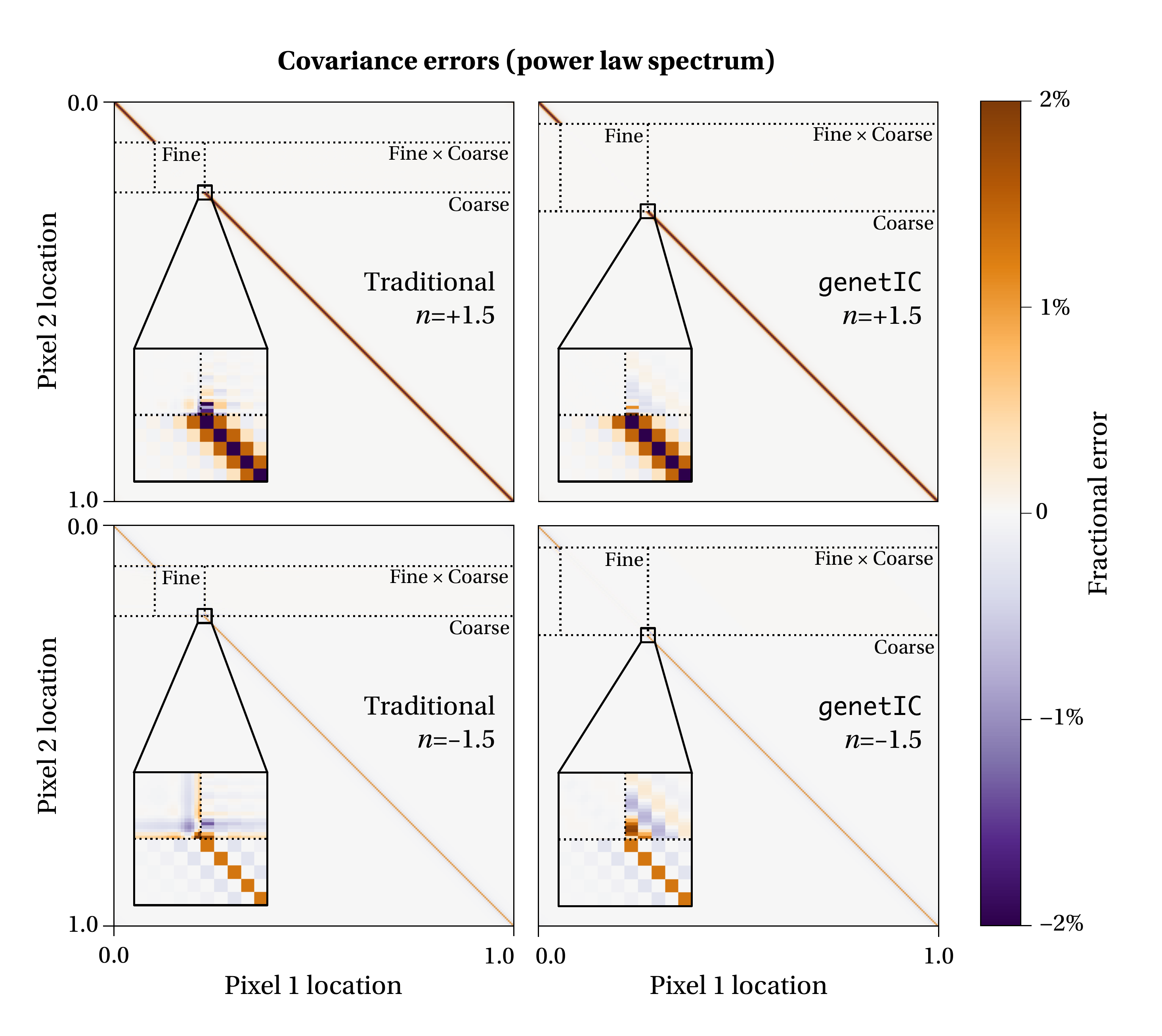}
	\end{center}
	\caption{Errors on the covariance matrix of zoom initial conditions for a 1D toy example of genetic modification. The base grid here is defined on a line between $0$ and $1$, and each panel shows the error in covariance between different pixels as a matrix. Two example 1D power spectra, $P(k)\propto k^{1.5}$ (top panels) and
		$P(k) \propto k^{-1.5}$ (lower panels) are tested. The errors for traditional (left panels) and Fourier filtering (right panels) approaches
		are shown relative to the idealized covariance matrix obtained by making
		realizations at high resolution across the whole box, then degrading
		the low resolution region.  We have fixed the computational demands, meaning that only half the high-resolution box is available for use in a traditional approach (the remainder being required as padding). The inset panels show the covariance error in the region around the transition from low-resolution to high-resolution pixelization. In both cases the largest errors are found throughout the low-resolution box due to power above the band limit. This error is present and unavoidable in all cosmological simulations. All other errors are small (of order $1\%$), making either approach acceptable for practical application to galaxy formation simulations. }
	\label{fig:cov-errors-trad-1}
\end{figure*}

To balance these requirements \codefont{genetIC} uses a Fourier-space Fermi-Dirac distribution illustrated in Figure \ref{fig:fermi-filter}:
\begin{equation}
\matrixfont{F}_L(k) = (\exp[(k-k_0)/(k_0T)]+1)^{-1}.\label{eq:fermi}
\end{equation}
The high resolution filter, $\matrixfont{F}_H$,  is then related to this by
\begin{equation}
\matrixfont{F}_L^2 + \matrixfont{F}_H^2 = \mathbb{I},\label{eq:filter-completeness}
\end{equation}
which is required in order to recover $\vdelta$ from the sum of the filtered fields $\matrixfont{F}_L \tilde{\vdelta}_L + \matrixfont{F}_H \tilde{\vdelta}_H$ or, equivalently, to obtain the correct power spectrum in the high resolution region (see Appendix \ref{app:derivation_fast_filter}). The wavelength at which the cutoff in the filter must occur should be larger than the Nyquist length, but smaller than the zoom region length-scale.  There will inevitably be some trade-off in the choice of the ``temperature'', $T$; the filter must remain sufficiently smooth in Fourier space that oscillations in the real space filter die away rapidly; but an excessively smooth filter will attempt to retain information at or above the Nyquist frequency leading to inaccurate small-scale correlations; see Figure~\ref{fig:fermi-filter}.

The default code choices are $T = 0.1$ and $k_0 = 0.5\,k_{\mathrm{nyq}}$, where $k_{\mathrm{nyq}}$ is the Nyquist frequency of the low-resolution grid. These were selected based on experimentation with the practical performance of the algorithm in a 1D test setting where exact covariances can be computed. The cosmological power spectrum is red on small scales ($k>0.2 h\mathrm{\,Mpc}^{-1}$) but blue on large scales ($k<0.2 h\mathrm{\,Mpc}^{-1}$); therefore we tested a range of power laws and the actual dimensionless cosmological spectrum (i.e. maintaining the total power per unit log wavenumber when moving from the 3D to the 1D setting). With our choice above, we found that fractional errors on covariances of the final field are always smaller than $\sim 2\,\%$ (Figure \ref{fig:cov-errors-trad-1}). In Figure \ref{fig:cov-errors-trad-1}, we compare with the ``traditional'' approach, which involves using a large padding region around the outside of the high-resolution box (resulting in a smaller usable high-resolution volume for the same computational effort compared to \codefont{genetIC}).

The dominant error term is aliasing on the low resolution grid, which reflects the existence of power above the Nyquist limit. This error is irreducible and present in {\it all} cosmological simulations, although the reddening of the spectrum at high\nobreakdash-$k$ means it rapidly becomes smaller at higher resolutions. Other errors are comparable to (or smaller than) this dominant term, and similar in magnitude between the Fourier space and traditional approaches. For this reason, we have not explored alternative filtering methods, although we note that (should increased accuracy ever be required) a multi-resolution wavelet \citep[e.g.][]{daubechies1992ten} approach to combining information from different zoom levels may be worth investigating.

\section{Modifications for Zoom Simulations}\label{sec:constraints}

The most important feature of \codefont{genetIC} is the generation of modified initial conditions. Fundamentally, we wish to map an existing field, $\nvec{\delta}$, on to a new field, $\nvec{\delta}'$ that satisfies a chosen condition, but is otherwise minimally altered. If the density field is stored at a single resolution for the whole simulation, and the condition is described by a linear function of the density field, then the solution is provided by the Hoffman-Ribak algorithm \citep{hoffman1991constrained}. 

Typically in Hoffman-Ribak applications the original, unconstrained field is treated as an intermediate object and not used for any computation. In modifications on the other hand, the unconstrained field is treated as physically meaningful; we must simulate the unmodified galaxy as well as a series of modified versions to probe how galaxies react to their surroundings or cosmological history. For modifications described by linear maps of the original density field, the resulting procedure is otherwise equivalent to the Hoffman-Ribak procedure. However, non-linear modifications result in a different procedure \citep{2018MNRAS.474...45R}.

In the case of a zoom simulation, it is not immediately clear how one should make modifications that are consistent across the boundaries between different resolution grids. We will next discuss broadly how this problem is solved in \codefont{genetIC}, and give full details of the algorithm and its derivation in Appendix \ref{app:constraints}.

\subsection{Linear Modifications}
\label{sec:linear}
	
To provide the starting point for modifications in zoom simulations, let us first describe more fully the approach for a uniform-resolution grid. In the simplest possible scenario, we want to change the average value of the field over some region, either multiplying it by a constant or setting it to a given value. Supposing the region corresponds to $N$ elements of the density field, $\delta_{i_1},\delta_{i_2}\,\ldots, \delta_{i_N}$, the constraint can be expressed as
\begin{equation}
\frac{1}{N}\left(\delta_{i_1} + \delta_{i_2} + \ldots + \delta_{i_N}\right) = \bar{\delta},\label{eq:localisation-covector}
\end{equation}
where $\bar{\delta}$ is the target average density contrast we wish to impose. Because such modifications are defined in terms of a linear sum of elements of the density contrast vector, $\nvec{\delta}$, we refer to them as \emph{linear modifications}. Because the potential and velocity fields are themselves linearly related to the density, any linear modification can be expressed as a constraint on the density field. In the general case, the constraint would be described by a vector $\nvec{u}$ and the target value $d$, and we aim to achieve
\begin{equation}
\nvec{u}\cdot\nvec{\delta}' = d.\label{eq:linearConstraint}
\end{equation}
In the average-density example, $\nvec{u}$ is a vector with components $1/N$ for elements of $\nvec{\delta}'$ that lie in the region to be modified, and zero outside it, and $d=\bar{\delta}$. 

The Hoffman-Ribak algorithm uses $\nvec{u}$ and $d$ to create a map from the unmodified field, $\nvec{\delta}$ to the modified field $\nvec{\delta'}$,
\begin{equation}
\nvec{\delta}' = \nvec{\delta} + \frac{(d - \nvec{u}\cdot\nvec{\delta})\matrixfont{C}\nvec{u}}{\nvec{u}\cdot\matrixfont{C}\nvec{u}}.\label{eq:HRalgorithm}
\end{equation}
 One can verify that $\nvec{\delta}'$ satisfies equation (\ref{eq:linearConstraint}) by applying the dot product with $\nvec{u}$ to both sides. However, this is not the only way of satisfying the constraint --- equation (\ref{eq:HRalgorithm}) is a special choice because it can be regarded as finding the solution $\nvec{\delta}'$ that minimizes $(\nvec{\delta}' - \nvec{\delta})^{\dagger}\matrixfont{C}^{-1}(\nvec{\delta}' - \nvec{\delta})$ subject to the constraint of equation (\ref{eq:linearConstraint}).  Consequently, modifications made this way are \emph{minimal}, and are the most likely way of satisfying the constraint that could have arisen from a Gaussian random field with correlation matrix $\matrixfont{C}$. This disfavors, for example, unphysical modifications such as sharp discontinuities in the density field. See \citet{Roth:2015wha,2018MNRAS.474...45R} for further discussion.
 
The method can be used to force extreme modifications to the field; for example, two neighboring individual pixels might be set by the user to take wildly different values, violating the expected continuity. In this case, the solution to Eq.~\eqref{eq:HRalgorithm} will be equally extreme: for example, a power spectrum estimate based on the single resulting field may differ significantly from the $\Lambda$CDM ensemble mean, and non-Gaussianity null tests may fail.   Such rare deviations are always present in any Gaussian ensemble, but have a large $\chi^2$ reflecting their rarity.  \codefont{GenetIC} helps users understand whether they are generating outliers by outputting the $\chi^2$ change between the modified and unmodified fields. Provided the change is of order unity or less, the new realization can be considered a similarly likely draw from the Gaussian random ensemble as the original unmodified field.

The procedure leading to Eq.~\eqref{eq:HRalgorithm} is suitable for a simulation with a uniform resolution over the whole simulation grid. We now explain how the procedure generalizes to the case of multiple resolutions combined using the filters from Section \ref{sec:fast_filter}. We may assume that the modification is specified on the highest resolution grid, since by construction this grid will contain the galaxy that we wish to alter. However, it is incorrect to apply the modification only on this high-resolution grid.

To understand why, consider a modification that changes the average value in some region $\Gamma$ which lies entirely within the highest-resolution grid.  First note that $\Gamma$, the set of particles which define a modification, is not the same as the set of particles that will be modified by it. We might imagine changing only the field within this region in order to match the constraint, but doing so would either be impossible in the overdensity due to mass conservation or, for any other quantity, would produce a discontinuity at the edge of $\Gamma$. Accordingly, it cannot  be a \emph{minimal} modification: a field with sharp discontinuities is highly unlikely to arise from a random draw from a Gaussian distribution.

As discussed above, one of the outcomes achieved by the Hoffman-Ribak algorithm is that the modification avoids such unphysical realizations by choosing a form that maximizes the likelihood for the modification to have been drawn from a Gaussian distribution. However, the result is that the field outside of the initially specified region $\Gamma$ is also modified, and in particular the necessary changes in the field $\nvec{\delta}$ will propagate out to regions outside the highest-resolution grid even if the set of particles that defined the modification lie entirely within it. If we only modify the highest-resolution grid, then there will be a discontinuity at the grid's edge which again would be unphysical or, at the very least, highly unlikely within the $\Lambda$CDM ensemble. Making the high-resolution grid sufficiently large that the modification is negligible outside of it is infeasible for the computational performance reasons discussed in Section~\ref{sec:zoom_review}. The modification must instead simultaneously alter the high and low-resolution grids in way that is self-consistent.

Our approach is to return to the idealized defining relations of $\tilde{\nvec{\delta}}_L$ and $\tilde{\nvec{\delta}}_H$, equation~\eqref{eq:FL} and~\eqref{eq:FH}, and their compressed versions $\nvec{\delta}_L$ and $\nvec{\delta}_H$. The defining relations can be used to deduce the required operations on $\nvec{\delta}_L$ and $\nvec{\delta}_H$ which are equivalent to applying the Hoffman-Ribak procedure to the underlying high-resolution field $\nvec{\delta}$. To achieve this, we concatenate $\nvec{\delta}_L$ and $\nvec{\delta}_H$ into a single compressed vector $\nvec{\delta}_Z$ ($Z$ here standing for ``zoom'').

Modifications form a map $\nvec{\delta} \to \nvec{\delta}'$ taking the underlying uncompressed density field into its modified counterpart. An ideal solution for the compressed modified field, $\nvec{\delta}_Z'$, would be obtained by first modifying the field $\nvec{\delta} \to \nvec{\delta}'$, then compressing $\nvec{\delta}' \to \nvec{\delta}_Z'$. For a practical algorithm, we are searching for a single operation that modifies $\nvec{\delta}_Z \to \nvec{\delta}_Z'$ in an equivalent way.  In other words, we want the operations of modification and compression to a zoom simulation to commute, at least approximately.

Given this principle, we can derive the required operation on the zoom vector $\nvec{\delta}_{Z}$ by regarding the operation of compression, mapping $\nvec{\delta} \rightarrow \nvec{\delta}_{Z}$ as a coordinate transformation\footnote{Strictly speaking this transformation is not invertible --- we address how to deal with this in Appendix \ref{app:constraints}.}. Then, the Hoffman-Ribak algorithm, whose application is straightforward for a fixed-resolution vector $\nvec{\delta}$, can be transformed into an operation defined in this ``zoom basis'' by the standard rules for performing coordinate transformations on linear operators. This results in a well-defined map from $\nvec{\delta}_{Z}$ to a new vector $\nvec{\delta}_{Z}'$ which satisfies the required constraint as if it had been performed at high-resolution everywhere and then degraded to the zoom simulation (to the level of precision that a given zoom set-up can describe). Having established this approach of demanding consistency with the original algorithm, the modifications can be applied at almost any stage of the initial conditions generation process  (see Section~\ref{sec:fast_filter}). In the final code, we choose to apply the modifications as early as possible, prior even to convolution with the transfer function.  The full details are derived and explained in Appendix \ref{app:constraints}.

\needspace{3em}
\subsection{Other Types of Linear Modification\label{sec:othermods}}

The prescription for modifying density contrasts is easily extended to modifying other quantities linearly related to the density contrast, such as velocity and the gravitational potential. For example the first order gravitational potential, $\Phi$, satisfies the Poisson equation which in Fourier space, assuming matter domination, is
\begin{align}
\Phi_k(\nvec{k}) =& -\frac{3\Omega_m H_0^2}{2a} \frac{\delta_k(\nvec{k})}{k^2},\label{eq:potential}
\end{align}
where the subscript $k$ denotes Fourier space quantities, $\Omega_m = \rho_m/\rho_{\mathrm{crit}}$ is the current matter density relative to the critical density, $H_0$ is the Hubble constant, and $a$ the scale factor when the initial conditions are generated.
Because the potential perturbation is linearly related to the density contrast, we can represent a constraint on the potential as a different constraint on the density contrast in the case of uniform resolution. One starts by constructing the  vector describing the region in which the potential is to be modified, just as in the density case; see equation~\eqref{eq:localisation-covector}. Next, the Poisson equation \eqref{eq:potential} is applied to $\nvec{u}$, allowing the Hoffman-Ribak algorithm to modify the potential while still operating on the overdensity field \citep{vanDeWeygaert96}. In the case of zoom simulations, one cannot perform this operation directly since solving the Poisson equation implies a convolution that leaks information from the high-resolution into the low-resolution region; instead we again use the uniform-resolution case as an idealized limit and so derive the correct procedure for zooms as described above in Section~\ref{sec:linear}.

Velocity modifications are implemented starting from the linear relationship between velocity and density contrast in the Zel'dovich approximation. Specifically, velocity perturbations are proportional to the gradient of the potential; during matter domination the  Fourier space relationship is given by
\begin{equation}
v_j(\nvec{k}) = -i \Omega_m^{1/2} H_0  a  \frac{k_j \delta_k(\nvec{k})}{k^2} .\label{eq:vel}
\end{equation}
Once more the uniform-resolution algorithm is obtained by applying the velocity transformation to the \nvec{u} vector; the variable-resolution case is again derived by requiring agreement with the idealized limit.

\subsection{Quadratic Modifications}
\label{sec:quad}
All of the modifications we have discussed up to this point are linear. Another type of modification consists of quadratic modifications to the field. These were first discussed by \citet{2018MNRAS.474...45R} and amount to modifying the density contrast vectors to satisfy
\begin{equation}
	\nvec{\delta}^{\dagger'}\matrixfont{Q}\nvec{\delta}' = q,\label{eq:quadMod}
\end{equation}
where $\matrixfont{Q}$ is a matrix. The simplest example of such a modification is altering the variance of an arbitrary region within the simulated density field. This can be useful, for example, in altering the smoothness of an overall halo merger history. The solution for uniform-resolution fields is already described in \citet{2018MNRAS.474...45R}, and consists of a gradient descent approximation involving a local linearization of the problem. Because each step in the gradient descent is approximated by a Hoffman-Ribak update, no extra work is required to implement the algorithm in zoom simulations. \codefont{genetIC} allows variance modifications to be requested by the user, and internally transforms these into an appropriate sequence of linear modifications. By constructing an appropriate $\matrixfont{Q}$ operator other types of quadratic modifications could also be computed in future, and \codefont{genetIC} is modularized to allow for these to be added easily.

\section{Accuracy and Examples}
\label{sec:examples}
\begin{figure*}
	\centering
	\includegraphics[width=0.8\textwidth]{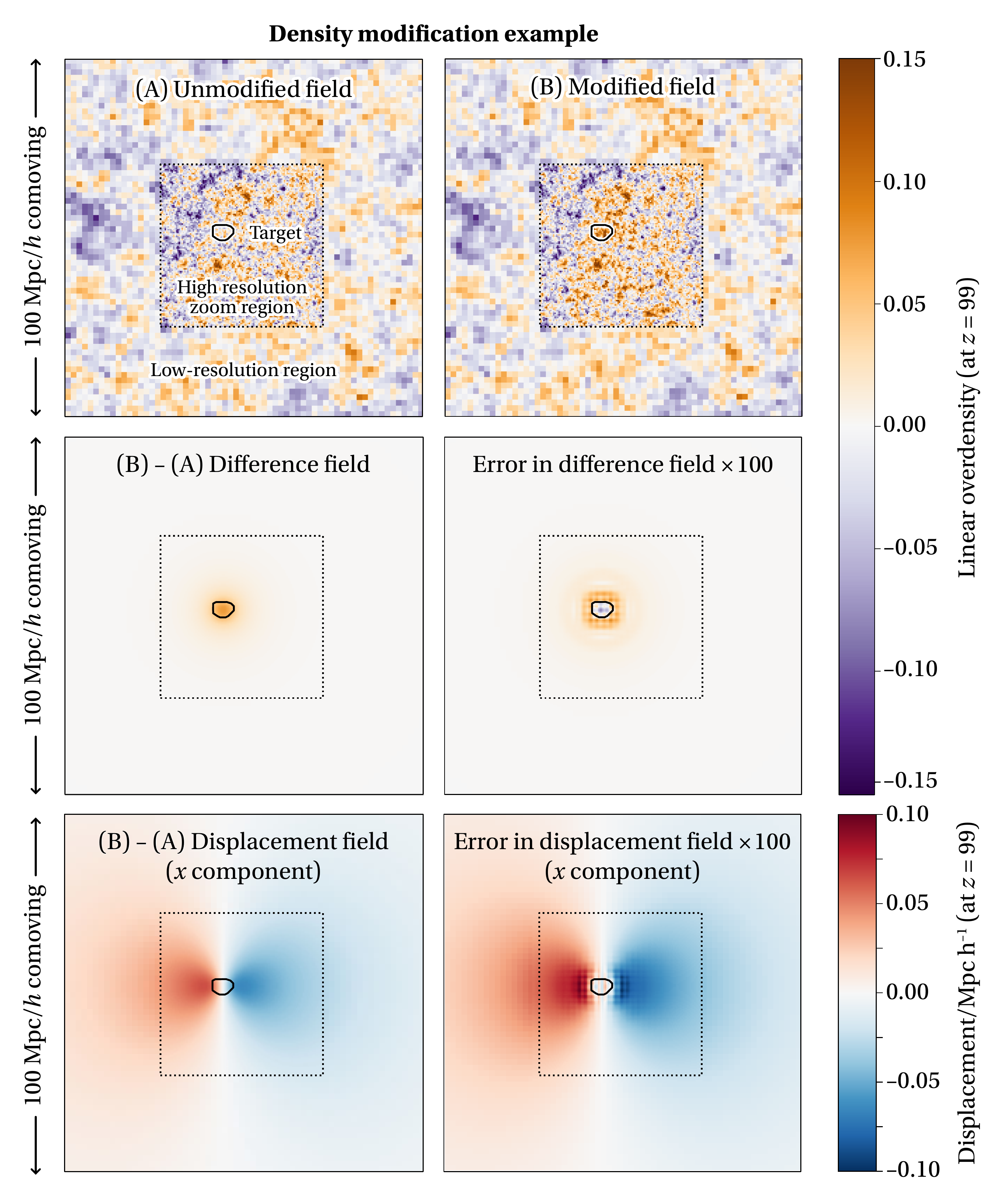}
	\caption{\label{fig:comparison}Density slices illustrating the modification of a halo's Lagrangian volume in a zoom simulation ($128^3$, $50\,\mathrm{Mpc}/h$ high-resolution grid embedded within a subvolume of a $128^3$ low-resolution grid which is $200\,\mathrm{Mpc}/h$ on a side; here only the central $100\,\mathrm{Mpc}/h$ region is shown). We change the average density contrast in the region indicated by the loop, while the dotted square represents the boundary of the high-resolution region. The middle left plot shows the difference between modified and unmodified fields, which is smooth across the resolution boundary (for a clearer example of this continuity see Figure~\ref{fig:velMod}). In the middle right panel we compare the difference field to that which is obtained in the idealized case where  the whole volume is realized at high-resolution, plotting the difference defined in equation (\ref{eq:Delta}) exaggerated by a factor of 100. We obtain percent level accuracy relative to the idealized case. The bottom panels show the same as the middle panels, but for the $x$ component of the displacement field, instead of the density field.}
\end{figure*}

To illustrate the accuracy of \genetIC~in performing convolutions across resolution boundaries, we will now verify the accuracy of its output when modifying zoom initial conditions. This can be accomplished by comparing to an ideal equivalent set of initial conditions which are realized at uniformly high resolution (and then degraded outside the zoom region). Studying the accuracy of these modifications complements our earlier investigation of the real-space correlation function generated when using the code's Fourier-split approach (Figure~\ref{fig:cov-errors-trad-1}.)\\

\subsection{Accuracy of the Initial Conditions}

\label{subsec:ic_accuracy}

The set up is as follows: we generate a $200 \mathrm{\,Mpc}/h$ cubic box at $z=99$ and  resolution $N_{\mathrm{sim}} = 512^3$, as the `high-resolution everywhere' initial conditions. These play the role of the ``ideal'' case, which for higher resolution zoom simulations would be infeasible to run --- at this relatively low resolution however, we can compare the ``ideal'' and zoom simulations directly. The zoom simulation is defined on the same size of box, but with a $128^3$ low-resolution grid and a zoom window consisting of a $50 \mathrm{\,Mpc}/h$ cube of resolution $N_{\mathrm{window}} = 128^3$ centered on $\nvec{x} = (50,50,50)\mathrm{\,Mpc}/h$. Note that its spatial resolution therefore matches that of the original $512^3$ box. Both sets of initial conditions are seeded with the same random seed in Fourier space. Because the low resolution modes are seeded first, they match between the simulations; however we caution that there is no way to make the high resolution modes match exactly in this test because the Fourier modes have different meanings between the idealized and the zoom case. This will necessitate a scaling in the comparison, which we will describe in due course.

\begin{figure*}
	\centering
	\includegraphics[width=0.8\textwidth]{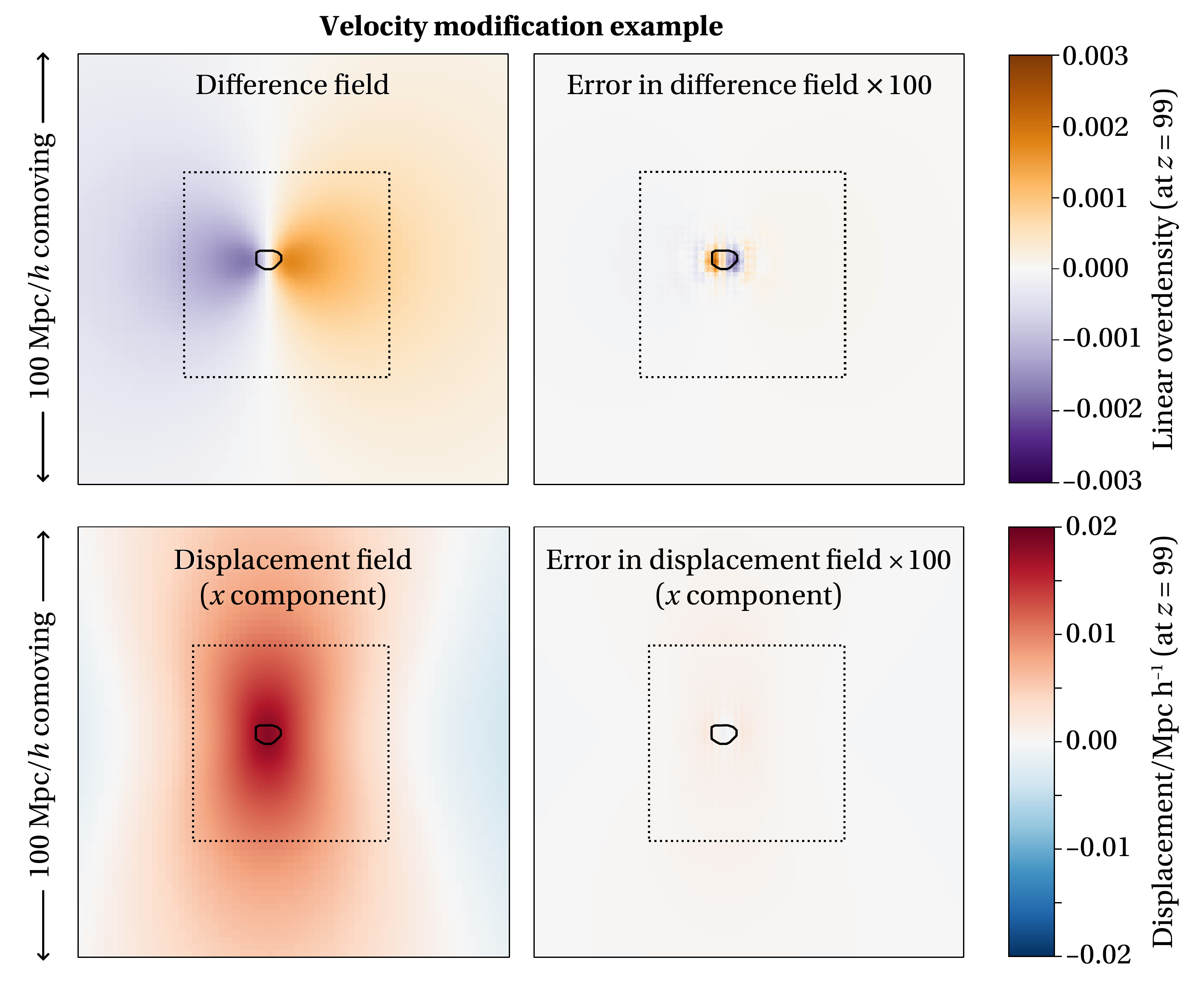}
	\caption{\label{fig:velMod} Top left: difference between modified and unmodified density contrast field after a velocity modification on the same grid as Figure~\ref{fig:comparison} ($128^3$ low-resolution $200\,\mathrm{Mpc}/h$ grid, $128^3$ high-resolution $50\,\mathrm{Mpc}/h$ region, indicated by the dotted square). As with density contrast modifications, the difference between modified and unmodified fields (left) is smooth across the boundary of the grids; this is now more clearly demonstrated because velocities are correlated on larger scales. Top right: error in this difference compared with the same modification on a higher, fixed resolution grid ($512^3$ across the full $200\,\mathrm{Mpc}/h$ box), exaggerated by a factor of 100. This exaggeration highlights a small discontinuity at the boundary between different resolution grids, but we emphasize that the error remains below percent-level. Bottom left: the changes to the displacement field (which in the Zel'dovich approximation is directly proportional to the velocity). Bottom right: the error in these changes compared with the same modification on the higher, fixed resolution grid, exaggerated by a factor of 100. The magnitude of the errors in this case are too small to be visible.}
\end{figure*}

An example slice through initial conditions on this grid is shown in Figure \ref{fig:comparison} (top left panel). The set of particles to be modified, $\Gamma$, is the Lagrangian region of a particular halo, chosen by evolving forward the unmodified zoom initial conditions, selecting a halo in the zoom window, and tracking its particles back to the initial conditions (see \citet{Roth:2015wha} and Section~\ref{sec:tracing} for a description of this process, which is unique to \codefont{genetIC}). A density contrast modification then makes minimal alterations to the field to change the density contrast averaged over those particles. For the purposes of this example we will choose the constraint on the modified field, $\nvec{\delta}'$, to be

\begin{equation}
\langle\nvec{\delta}'\rangle_{\Gamma} = 0.1\label{eq:egConstraint}.
\end{equation}

This describes a particularly large modification, comparable to the r.m.s. of the entire field, for the purposes of illustration. The effect on the zoom initial conditions is shown in Figure \ref{fig:comparison} (top right panel). In the middle left panel, we show the difference between the modified and unmodified fields (with the same color scale as the overdensity field itself). To characterize the error in this modification relative to the ``ideal'' case, we subtract the difference between the zoom simulation modification and a rescaled ideal modification
\begin{equation}
\Delta(\vec{x}) = \left[\delta'_{\mathrm{Zoom}}(\vec{x}) - \delta_{\mathrm{Zoom}}(\vec{x})\right] - \alpha\left[\delta'_{\mathrm{Ideal}}(\vec{x}) -\delta_{\mathrm{Ideal}}(\vec{x})\right],\label{eq:Delta}
\end{equation}
where $\delta_{\mathrm{Ideal}}(\vec{x})$ and $\delta_{\mathrm{Zoom}}(\vec{x})$ represent the field as computed in the idealized and zoom cases respectively, and primes indicate modified fields.  The factor $\alpha$ rescales the ideal modification field. This is needed to make a proper comparison because the pre-modification average of the density contrast field in any given region is not the same for idealized and zoom fields, that is, $\langle\nvec{\delta}_{\mathrm{Zoom}}\rangle_{\Gamma} \neq \langle\nvec{\delta}_{\mathrm{Ideal}}\rangle_{\Gamma}$: the fields unavoidably have different high resolution modes simply due to the way that random noise is seeded. Consequently, the modification fields $\tilde{\nvec{\delta}}_{\mathrm{Ideal}} - \nvec{\delta}_{\mathrm{Ideal}}$ and $\tilde{\nvec{\delta}}_{\mathrm{Zoom}} - \nvec{\delta}_{\mathrm{Zoom}}$ that enforce equation (\ref{eq:egConstraint}) have different amplitudes. The rescaling factor needed to compare them is just the ratio of these modification amplitudes
\begin{equation}
\alpha = \frac{\langle\tilde{\nvec{\delta}}_{\mathrm{Zoom}}\rangle_{\Gamma} - \langle\nvec{\delta}_{\mathrm{Zoom}}\rangle_{\Gamma}}{\langle\tilde{\nvec{\delta}}_{\mathrm{Ideal}}\rangle_{\Gamma} - \langle\nvec{\delta}_{\mathrm{Ideal}}\rangle_{\Gamma}}.
\end{equation}

\begin{figure*}
	\centering
	\includegraphics[width=0.9\textwidth]{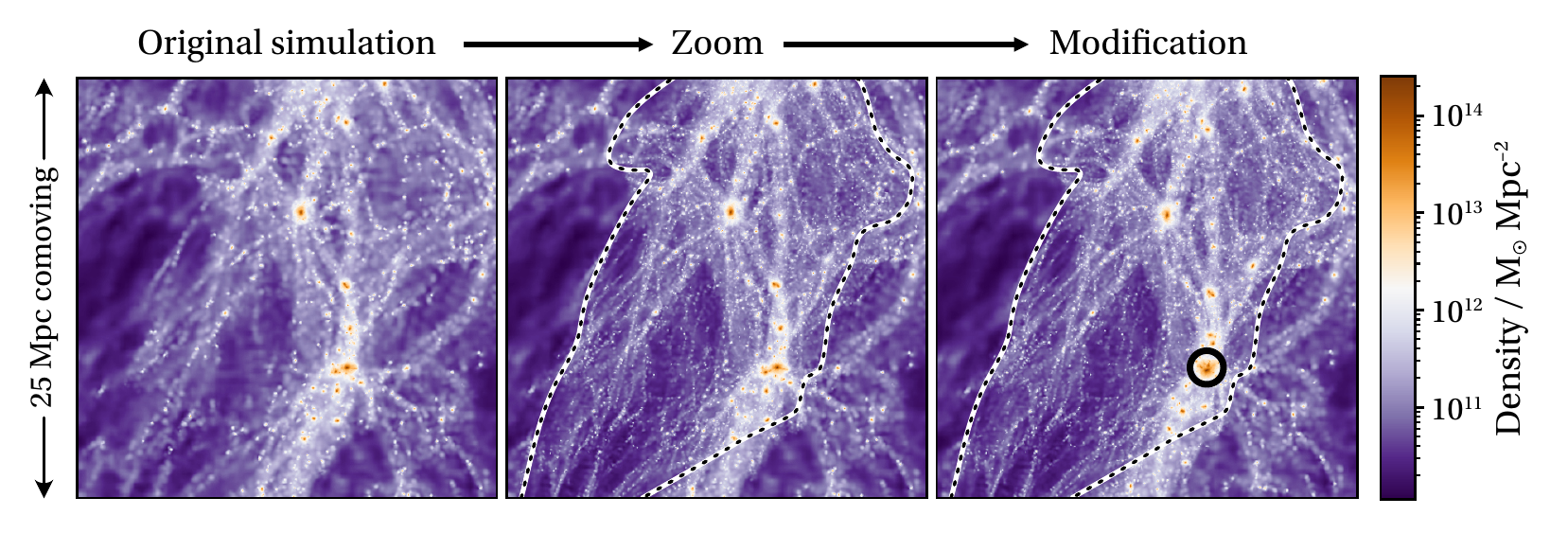}
	\caption{\label{fig:evolved} An example of {\genetIC} in use. Left: density in a $5$\,Mpc-thick slice through a $100\,\Mpc/h$ uniform-resolution simulation, evolved to $z=0$ using {\tt GADGET2}. Center: the same density slice taken through a simulation in which we have created a large cubic `zoom' region with 8 times better mass resolution. (The zoom region deforms away from a cube due to non-linear evolution, and its boundaries are shown as a dashed line.)  Large-scale structures persist between the unzoomed and zoomed simulation, and increased detail can be seen in the zoomed version.   Right: a halo (circled) within the zoom region is selected and genetically modified to increase its initial density by $20\%$, and the simulation is performed once again. The result is to increase the halo's $z=0$ mass from $8.73\times 10^{12}\,\Msol/h$ to $1.78\times 10^{13}$\,$\Msol/h$ (over a factor of two increase) while making minimal changes to the surrounding structure. Note that due to small movements perpendicular to the page, small halos can move in and out of our slice, so spuriously appear or disappear in the right panel.}
\end{figure*}

The final result of computing the error, equation~\eqref{eq:Delta}, is plotted in the middle right panel of Figure \ref{fig:comparison}, exaggerated in scale by a factor of $100$ relative to the fields and modifications themselves, showing that the error in modifications is vastly smaller than the modifications themselves. Since modifications in practical scenarios will generally be much smaller than the example presented here, this small error on already small modifications will be  negligible.  Crucially, the modification field (defined as the difference between the modified and unmodified fields) is smooth across both the boundary of the set of particles defining the modification and the grid boundaries (middle left panel of Figure~\ref{fig:comparison}). We also show, in the bottom panels of Figure \ref{fig:comparison}, the difference between the modified and unmodified displacement field ($x$ component, bottom left), and the error in the same displacement field magnified by a factor of 100 (bottom right). This illustrates the smoothness of the actual velocities and particle displacements that will affect the resulting particle distribution directly.

The continuity across the boundary is better seen by making a longer-wavelength, high-amplitude modification. In Figure \ref{fig:velMod}, we consider a modification to the velocity field, obtained by constraining the average over $\Gamma$ of the $x$-component of the velocity to be $100\,\mathrm{kms}^{-1}$. Because of the additional $k^{-1}$ weighting of equation~\eqref{eq:vel}, the corresponding modification affects wavelengths up to the fundamental mode of the box. (Velocity correlations in cosmological simulations extend up $\simeq 1\,\mathrm{Gpc}$ if the box size is sufficiently large.) The modification is accurately propagated outside the zoom window with only small errors compared to performing the same modification on a higher resolution grid (note that the errors in the top and bottom right panels are again magnified by a factor of one hundred --- although these highlight discontinuity in the errors, the overall modification field is still accurate and continuous to better than percent level).

\needspace{2cm}
\subsection{Accuracy of Evolved Examples}

\label{subsec:evolved_accuracy}

The discussion in Section \ref{subsec:ic_accuracy} demonstrates the accuracy of \genetIC's approach to convolutions which transfer information from small to large scales. However, it is also important to establish that large scale information is correctly propagated into zoom regions; of particular concern is the persistence of  structure when creating zoom simulations and modifying them. To demonstrate this persistence, we ran a $100\, \mathrm{Mpc\,}h^{-1}$, $512^3$, dark-matter only `base' simulation with $\Omega_m = 0.3156, h = 0.6726, \sigma_8 = 0.830$ \citep{Planck16}. We then generated a second set of `zoomed' initial conditions that introduced eight times higher mass resolution within a $20\,\mathrm{Mpc\,}h^{-1}$ cube at the center.  Because both simulations are generated with the same random seed, the low-resolution modes of both match. Both base and zoomed simulations were evolved with \codefont{GADGET2} \citep{Springel:2005mi}.

We show in Figure \ref{fig:evolved} the evolved state at $z = 0$ in the selected region; the left panel shows a projection of a $5\,\Mpc$ slice through the original simulation, and the center panel shows the zoomed simulation, with the dashed line showing the approximate boundary of the zoom region in projection. The large-scale structure and locations of resolved satellite halos are preserved, while additional small-scale structure is introduced as expected. The mass of individual resolved structures changes by less than $1\,\%$.

We then selected a halo in the zoom region, and increased the density contrast of its constituent particles in the initial conditions by $20\%$. This increased the mass of the final halo from $0.88\times 10^{13}\,\Msol/h$ to $1.78\times 10^{13}\,\Msol/h$, as expected \citep{Roth:2015wha}. The evolved, modified density field is shown in the right-most panel of Figure \ref{fig:evolved}. The modifications to the halo result in small changes to its surrounding structures, maintaining consistency with the $\Lambda$CDM power spectrum. These changes include small shifts in the location of structures, some of which are perpendicular to the projected slice (which means structures can disappear from the Figure).  However, the overall  large scale structure is maintained, as expected.

\section{Code Overview and Core Features}

\label{sec:overview}

We now give a technical overview of the \codefont{genetIC} code. There are three main stages involved in a typical scenario when using  \codefont{genetIC}: (1) grid creation and white noise generation; (2) modification; and (3) particle generation. These stages are illustrated in Figures \ref{fig:overview} and \ref{fig:generator}, together with the code classes involved listed underneath each step of the stages. We do not describe the user syntax for controlling these stages through a parameter file, since these are detailed in a separate manual. The code is implemented in C++, with parallelization via OpenMP. It relies on the GNU Scientific Library \citep{GoughGSL} for random number generation and FFTW for Fourier transforms \citep[e.g.][]{fftw}.

\subsection{Stage 1 -- Grid set-up and White Noise Generation}
\label{sec:stage1_setup}

\begin{figure*}
	\includegraphics[width=\textwidth]{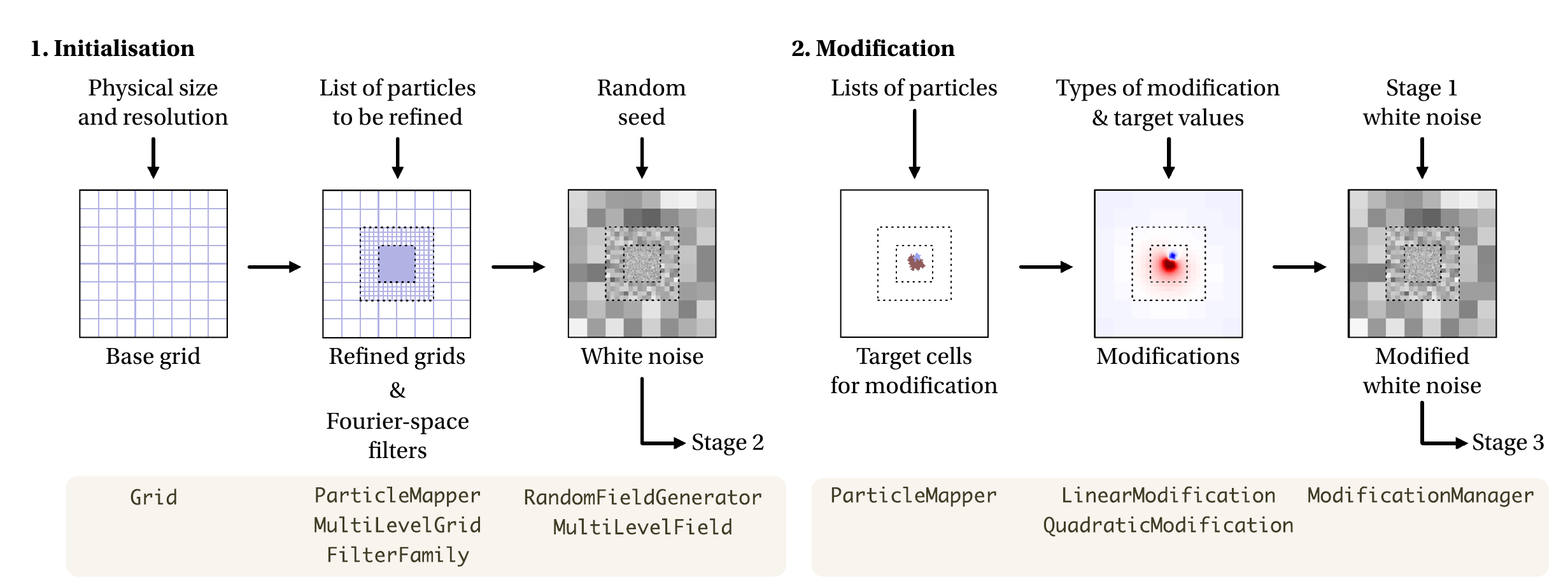}
	\caption{\label{fig:overview}To provide an overview of the initial conditions generation process within \codefont{genetIC}, we break it down into three stages; here the first two are illustrated. Control passes from left to right. Names of the classes involved at each step are given below each stage, and inputs flow from above. In step one,  information specified by the user in a parameter file is used to construct a set of grid objects. If zoom levels are required, refinement grids are constructed. White noise fields are then generated for each grid. The second step only occurs if modifications are required, using information in the parameter file to construct and apply these to the white noise. Step three is illustrated in Figure~\ref{fig:generator}.}
\end{figure*}

In the first stage (see Figure \ref{fig:overview}), the parameters are used to construct a \classname{Grid} object that stores properties of the simulation, such as the cell size, particle mass, and dimensions. To enable zoom simulations, refinement grids are set up according to the specification in the parameter file and organized into a \classname{MultiLevelGrid} class that encapsulates the relationship between different grids.  It stores data about each level's size, grid layout, and position, as well as various functions for accessing and manipulating these grids. Any part of the code needing access to the relationship between different grids uses this object.

At the same time, a \classname{ParticleMapper} object is set up to track how the grids are related to particles: for example, if the user specifies that a sphere around some point of the simulation should be stored at higher resolution, then a grid is set up that contains this sphere and the \classname{ParticleMapper} keeps track of which particles in that sphere were requested and should be included at the particle generation stage (see Section \ref{sec:stage3_generation}). The \classname{ParticleMapper} is able to map bi-directionally between grid cells to particles, an essential facility in the modification stage below. Performing this mapping can be expensive in complicated geometries, but is parallelized for maximum efficiency.

Once the grids are initialized, the code constructs a \classname{MultiLevelField} object that contains \classname{Field} objects, each of which stores the overdensity field on a single \classname{Grid}. A \classname{RandomFieldGenerator} uses the random seed specified in the parameter file to generate unit variance white noise within each \classname{Field}. The \classname{RandomFieldGenerator} makes use of GNU Scientific Library's Ziggurat algorithm \citep{GoughGSL} to draw uncorrelated Gaussian samples, starting from a user supplied seed integer (this allows the same set of random numbers to be drawn, irrespective of system architecture). Separating this module into its own object allows the random number generator back-end to be easily changed, and there are several options for how the random components of the  fields can be generated. In particular, the random draws can be performed both in Fourier and real-space, and either in series or in parallel according to the user's requirements. When drawing in parallel, the work is segmented into Fourier-space shells which can be drawn independently, so that the final result is independent of the number of threads available.\\

Note that it is also possible to import pre-existing white noise fields instead of drawing random noise, such as those that might have been generated by other initial conditions generators. These must be in \codefont{numpy} format, and be of the correct size to fit the grids to be generated. This allows the possibility of applying genetic modifications to existing simulation suites.

\needspace{2cm}
\subsection{Stage 2 -- Modifications}
\label{sec:stage2_modification}
In the second stage (see Figure \ref{fig:overview}), modifications are applied. These are specified in the parameter file and can be a linear or quadratic function of a set of particles (such as the average density). The \classname{ParticleMapper} is used to trace the target particles into appropriate cells within the \classname{MultiLevelGrid} structure. The specified modifications are used to construct \classname{LinearModification} and \classname{QuadraticModification} objects as appropriate, which are stored by the \classname{ModificationManager} --- this object is the heart of the \codefont{genetIC} algorithm. It implements all modification algorithms, and applies them to the white noise field. 

The linear modifications currently implemented are density, potential, and velocity in three Cartesian directions. They are each represented by \classname{LinearModification} subclasses, thus allowing future expansion or custom modifications to be defined. For quadratic modifications, only variance is currently implemented, and is in a subclass of \classname{QuadraticModification} so that future expansion should be straightforward.

\subsection{Stage 3 -- Particle Generation}
\label{sec:stage3_generation}

\begin{figure*}
	\centering
	\includegraphics[width=\textwidth]{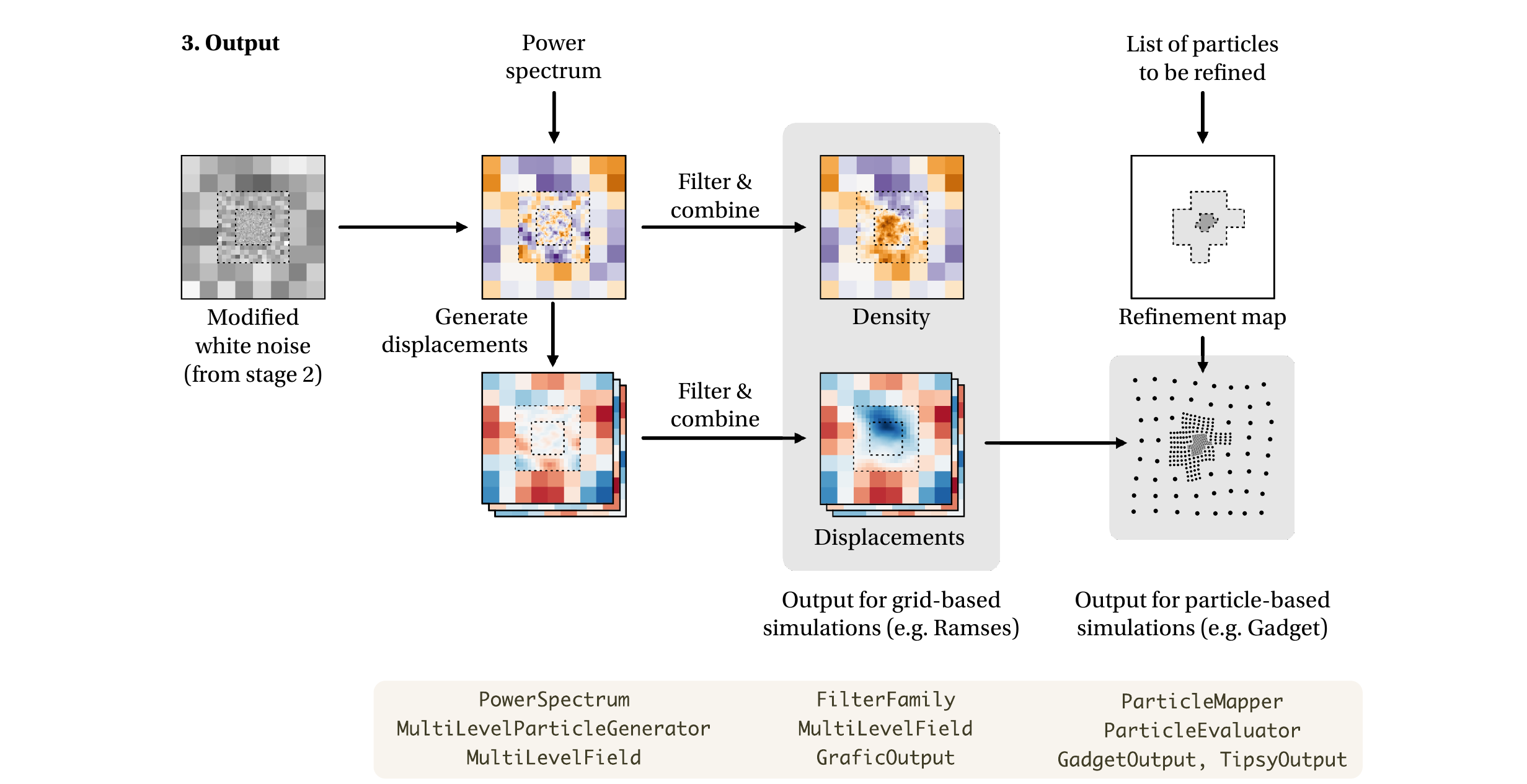}
	\caption{\label{fig:generator}Stage three of \codefont{genetIC}. As in Figure \ref{fig:overview}, classes for each step are indicated below, and inputs above. The (modified) white noise on each level has the power spectrum applied in Fourier  space and is used to construct displacement fields for the particles. At this stage, the short-wavelength behavior of zoom grids is incorrect, i.e. the zoom regions do not match the long wavelength modes of the base grid. This is corrected by filtering and combining the displacement fields on each level to give the correct long and short wavelength behavior in all zoom regions. Finally, the displacement fields are converted into particles or gridded output, ready for use with a simulation code.}
\end{figure*}

In the final stage, the modified white noise fields are converted into particle positions and velocities. This step is illustrated in Figure \ref{fig:generator}. First, the power spectrum is applied to the white noise field on each grid; however, this does not result in the final overdensity since the filtering and mode combination step (Section~\ref{sec:fast_filter} and Appendix~\ref{app:derivation_fast_filter}) has not yet been undertaken. The \classname{MultiLevelParticleGenerator} object takes responsibility for the required steps, as well as producing displacement and velocity fields which are required for the simulation initialization. The latter are generated by applying the appropriate convolution, such as equation~\eqref{eq:vel}, on the individual grids independently, and then filtering and combining the modes in the fields just as for the overdensity field.  While currently the code assumes the Zel'dovich approximation during these manipulations, other methods, such as higher order Lagrangian perturbation theory, could be implemented in future by adding subclasses.

Finally the code must map the fields as stored on the grids onto output particles. There are two interrelated aspects of this: (1) the correspondence between particles and grid cells of the field, and (2) the selection of a resolution for each region. To output particles taking account of these considerations requires bringing together information in the \classname{MultiLevelParticleGenerator} and the \classname{ParticleMapper}. For example, consider selecting a sphere of a given radius around a specific point (or a more abstract region such as the particles surrounding a particular halo). If we want to output these particles at high resolution,  the code internally creates a cubic grid around them and generates a white noise field for the whole cube. However, for efficiency in the final simulation we only want to output those particles explicitly selected for high resolution. We might also wish to insert a thin layer of ``intermediary'' particles which interpolate between the high and low resolution, or embed further zoom grids hierarchically. The \classname{ParticleMapper} keeps track of this information and determines for each point whether we should output a low-resolution particle or a group of high-resolution particles, allowing \codefont{genetIC} to handle detailed nested geometries seamlessly while shielding other parts of the code from these complexities.

Additionally, a \classname{ParticleEvaluator} object takes responsibility for outputting the appropriate fields for each grid. Depending on the precise simulation being used, and the options specified, different derived classes of the \classname{ParticleEvaluator} are used. In particular, the code can generate particles at lower resolution than the underlying random fields (sub-sampling), or at higher resolution via interpolation (super-sampling). These facilities are used when generating intermediary regions (see above) and are also available for explicit invocation by advanced users who wish to fine-tune the performance of a simulation. Further details can be found in the code and its documentation.
	
In summary, the \classname{MultiLevelParticleGenerator} takes the field and turns it into the raw information for particles, by default using the Zel'dovich approximation.  The \classname{ParticleMapper} keeps track of particles to be generated as output, while the \classname{ParticleEvaluator} keeps track of how the particles are related to the underlying density fields. To generate output, the \classname{ParticleMapper} iterates over the particles to be generated, accessing the relevant density fields via an intermediary \classname{ParticleEvaluator}. The mapper is also used to identify and trace cells corresponding to a list of particles from a previous output, as we now describe.

\needspace{3em}
\subsection{Tracing Particles}\label{sec:tracing}

A key feature of \codefont{genetIC} is its ability to map particles from a simulation back to an initial conditions grid cell. In some cases, this is a straight-forward operation: if we perform a single unzoomed simulation, the mapping from the final particle index to the original grid position is simple. This can be used, for example, to select which grid cells will be refined when constructing a zoom simulation. However, if we then wish to generate a third set of initial conditions, in which we modify a region of the zoom simulation, the mapping is now considerably more complex, since only a fraction of the high resolution grid is represented as particles. Additional complexities arise if baryon particles are added in a subregion; if a second zoom level is opened; if intermediate resolution `padding' particles are added around a high resolution region; or if genetic modifications change the set of particles which ultimately fall into the target galaxy. 

One simple way around the mapping problem is to require that the simulation code associates an ID with each particle, with the ID uniquely identifying the origin grid cell, regardless of which cells are represented in the particle output. However, some codes do not offer this option; moreover, this kind of non-contiguous particle labeling can pose significant challenges for simulation post-processing and analysis routines.

The \codefont{genetIC} code can instead map from sequential particle IDs back to the original grid squares, when used with Lagrangian codes such as \codefont{GADGET} and \codefont{ChaNGa}. The user is required to specify the original parameter file that details how the original simulation was set up. By creating a second \classname{ParticleMapper} object to process the input particles IDs, the code is then able to trace the relationship between particles and grid cells.

\subsection{Baryon Transfer Functions}\label{sec:baryon-tf}
Often in $N$-body simulations it is assumed that because dark matter is the dominant matter component, the difference between the baryon and dark matter transfer functions can be neglected, and the total matter transfer function used for both. At late times this is a good approximation \citep[see][]{1990ApJ...361..343L,peebles1980large}. However, at early times the baryon transfer function contains features not present in the dark matter power spectrum --- most importantly the effect of baryon acoustic oscillations. As cosmological parameters become increasingly precisely determined, there may be a need in the future to take this difference into account.\\

For this reason, \codefont{genetIC} includes an option to take into account both the baryon and dark matter transfer functions in generating the initial conditions. This is accomplished by taking a copy of the white noise field after modifications but before applying the transfer function, and thus requires additional memory. However, the use of separate transfer functions is optional, and the user must explicitly enable it; otherwise, by default, the initial conditions use only the dark matter transfer function even for gas particles.\\

\subsection{Paired and Fixed Initial Conditions}\label{sec:pair-fix}

It is also possible to generate paired and fixed initial conditions. Paired fields have opposite phase in Fourier space, and so some correlations cancel between them, reducing sample variance while retaining Gaussianity. The effect is to swap overdensities and underdensities which in itself may be useful for contrasting the growth of halos and voids \citep{PhysRevD.93.103519}. Fixed fields, on the other hand, set the power spectrum to the exact theoretical mean and thus destroy Gaussianity in a controlled way \citep{Angulo:2016hjd}; their properties are discussed further in \citet{Villaescusa-Navarro:2018bpd}.  

\needspace{6em}
\section{Discussion}\label{sec:disc}
We have presented \codefont{genetIC}, an initial conditions generator for cosmological simulations. The code generates multi-level zoomed initial conditions and is specifically designed to perform modifications to these initial conditions in a controlled manner. 

\codefont{GenetIC} uses a new Fourier-space filtering approach to combine information from different resolution regions of the initial conditions (Section~\ref{sec:desc}). This avoids the need for large ghost regions around each level of the simulation, enabling deeper and higher-resolution zooms, and also allows the code to self-consistently propagate information about modifications made on the highest-resolution grid up to the lower-resolution grids.  

The algorithms also by construction implement minimal modifications (Section~\ref{sec:constraints}) --- that is, modifications that satisfy given constraints while minimizing changes and maximizing the likelihood of a given realization to have arisen from a Gaussian distribution. This prevents modifications from possessing unphysical features, such as sharp discontinuities at the boundary of regions that are modified. We verified that our implementation of multi-resolution initial conditions produces correlation functions and modifications that are in close agreement with idealized uniform high-resolution equivalents (Section~\ref{sec:examples}).

The code implements these concepts in an efficient parallelized manner (Section~\ref{sec:overview}). In order to enable the specification of complex geometries for zoom regions and modifications, it has a sophisticated mapping system that ingests lists of particle IDs from prior simulations and identifies the associated grid cells for further manipulation (Section~\ref{sec:stage1_setup}).  The code has a variety of useful additional features, such as generating gas with the correct baryon transfer function (Section~\ref{sec:baryon-tf}) and generating reversed initial conditions for the same initial seed (Section~\ref{sec:pair-fix}). One of its strongest features, however, is its modular, object-oriented design that allows for easy extensibility to apply the code to different situations. A test suite accompanies the code to verify installations and ensure code quality. Support for the code and download links can be found at \url{https://github.com/pynbody/genetIC}. An extensive user manual is available from the github site, and the code is  documented with Doxygen.
\section*{Acknowledgments}

We gratefully acknowledge helpful comments and suggestions from the referee, Oliver Hahn. We would also like to thank Corentin Cadiou, Luisa Lucie-Smith, and Chris Pedersen for useful feedback. This project has received funding from the European Union's Horizon 2020 research and innovation programme under grant agreement No.  818085 GMGalaxies. SS and AP were supported by the Royal Society. SS, NR and HVP were partially supported by the European Research Council (ERC) under the European Community's Seventh Framework Programme (FP7/2007-2013)/ERC grant agreement number 306478-CosmicDawn.   HVP was additionally supported by the research project grant ``Fundamental Physics from Cosmological Surveys'' funded by the Swedish Research Council (VR) under Dnr 2017-04212. MR acknowledges support from the Perren Fund, the IMPACT fund and the Knut and Alice Wallenberg Foundation.  This work was performed in part at the Aspen Center for Physics, which is supported by National Science Foundation grant PHY-1607611. This work was also partially supported by a grant from the Simons Foundation, and by funding from the UCL Cosmoparticle Initiative.

\appendix

\section{Mathematical Description of the GenetIC Algorithm}
\label{app:technical}

In Sections~\ref{sec:fast_filter} and~\ref{sec:constraints} we qualitatively described the approach implemented by \codefont{genetIC} to obtain and manipulate initial conditions for zoom simulations. We now provide a full analytic description, including further justification and derivations where appropriate. 

We start by laying out some notation in Appendix~\ref{sec:pixelandwindow}. Appendix~\ref{app:variable-resolution-to-fixed} then describes the relationship between white noise that is ``compressed''  (i.e. realized at variable resolution) and a uniform-resolution density contrast field. Approximations required to build a practical algorithm are outlined in Appendix~\ref{sec:simplify}. Finally, Appendix~\ref{app:constraints} describes how we apply modifications. 
For completeness, Appendix~\ref{app:even-more-technical} provides further technical insight into approximations. 

As in the main paper, we assume throughout the Appendix that there is a single zoom region, allowing us to make the notation as clear as possible. The code in fact supports nested regions where the resolution increases on each successive level, but this general multi-level case follows from recursion on the two-level approach.  

\subsection{Notation: Pixelization and Windowing \label{sec:pixelandwindow}}

To express our construction of zoom initial conditions as accurately as possible, it is useful to introduce some notation. We will consider two key operations: pixelization and windowing. Conceptually, these both start from a field sampled at uniformly high resolution across the simulation domain, and respectively down-sample to low resolution or extract the zoom region. 

We start in each case from the vector $\nvec{\delta}$, which contains a list of $N$ field values for each pixel in the simulation when sampled at uniformly high resolution. In practice, we never generate fields in this way since the goal is to avoid storing or manipulating such a prohibitively large vector (Section~\ref{sec:ics_review}). However, the operators are nonetheless needed for describing and justifying the algorithm in the remainder of the Appendix.

 The first operation, pixelization, will be denoted by $\matrixfont{P}$.  It down-samples $\vdelta$ to a low-resolution vector $\nvec{\delta}_P$ of length $N_P$, i.e.,
\begin{equation}
\nvec{\delta}_{P} = \matrixfont{P}\nvec{\delta}.
\end{equation}
Explicitly, $\matrixfont{P}$ can be represented by an $N_P\times N$ rectangular matrix. The simplest resampling scheme is to create each low-resolution pixel by averaging over $m = N/N_P$ high-resolution pixels. For example, in the case of $m = 3$, we can visualize the $\matrixfont{P}$ matrix as   
\begin{equation}
\matrixfont{P} = \frac{1}{3}\left(\begin{matrix}1&1&1&0&0&0&0 & \cdots \\0&0&0&1&1&1&0&\cdots\\0&0&0&0&0&0&1 & \cdots \\ &\vdots&&&\vdots&&&\ddots \end{matrix}\right).\label{eq:pixel-example}
\end{equation}

The other key operation is windowing, which selects a subset of the $\nvec{\delta}$ vector corresponding to the region in which we wish to retain high resolution initial conditions:
\begin{equation}
\nvec{\delta}_{W} = \matrixfont{W}\nvec{\delta}.
\end{equation}
The matrix $\matrixfont{W}$ has dimensions $N_W \times N$. The simplest explicit example in this case is given when the pixels in the zoom region are already located  at the beginning of the original vector $\vdelta$ (which in fact we may assume without loss of generality). Taking $N_W=3$ for illustrative purposes, one would have 
\begin{equation}
\matrixfont{W} = \left(\begin{matrix}
	1 & 0 & 0 & 0 & \cdots \\ 
	0 & 1 & 0 & 0 & \cdots \\ 
	0 & 0 & 1 & 0 & \cdots 
\end{matrix}\right)\,.\label{eq:window-matrix}
\end{equation}

Pixelization and windowing both destroy information, and so cannot be inverted, but we will make use of their pseudo-inverses $\matrixfont{P}^+$ and $\matrixfont{W}^+$. These respectively up-sample low-resolution field vectors to high resolution (in a specific way to be described shortly), and place the zoom region back into a full-sized simulation using zero-padding. They are defined to satisfy 
\begin{equation}
	\matrixfont{P}\matrixfont{P}^{+}=\mathbb{I}_{N_P}\quad\textrm{and}\quad \matrixfont{W}\matrixfont{W}^{+}=\mathbb{I}_{N_W},
\end{equation} 
where $\mathbb{I}_n$ is the $n\times n$ identity matrix. In other words up-sampling then down-sampling, or zero-padding then removing the zero padding, must have no effect.

From this requirement, one may derive explicit expressions for $\matrixfont{P}^+$ and $\matrixfont{W}^+$. In the latter case, we have simply $\matrixfont{W}^+ = \matrixfont{W}^{\dagger}$ since $\matrixfont{WW}^{\dagger} = \mathbb{I}$; this is most easily verified by inspection of the example~\eqref{eq:window-matrix}, which generalizes to any $N_W$. 
 
Deriving $\matrixfont{P}^+$ takes a little more care; recall that the pixelization $\matrixfont{P}$ forms each low resolution pixel from averaging over $m$ high resolution pixels. The pseudo-inverse simply places that mean value back into each high resolution pixel, leading to the expression $\matrixfont{P}^+ = m \matrixfont{P}^\dagger$. One may verify by inspection that for the example given in~\eqref{eq:pixel-example}, $m \matrixfont{PP^{\dagger}} = \mathbb{I}_{N_P}$ as required; this generalizes to any value of $m$.

Finally we note that the operators $\matrixfont{P}^{+}\matrixfont{P}$ and $\matrixfont{W}^{+}\matrixfont{W}$, describe respectively downsampling then upsampling the field and extracting the zoom region then zero-padding. These are both destructive operations but satisfy the projection relations 
\begin{equation}
	\left(\matrixfont{P}^+\matrixfont{P}\right)^2=\left(\matrixfont{P}^+\matrixfont{P}\right)\quad\textrm{and}\quad \left(\matrixfont{W}^+\matrixfont{W}\right)^2=\left(\matrixfont{W}^+\matrixfont{W}\right)\,.
\end{equation} 
Thus, there is no additional effect from repeated up-sampling and down-sampling or windowing and zero-padding. All the above relations are used routinely in derivations. 

One can also construct operators that pixelize only in the window region, denoted $\matrixfont{P}_W$, or window the pixelized field, denoted $\matrixfont{W}_P$ (as well as their pseudo-inverses, $\matrixfont{P}_W^+$ and $\matrixfont{W}_P^+$). The order between pixelizing and windowing does not matter, so that
\begin{equation}
\matrixfont{P}_W\matrixfont{W} = \matrixfont{W}_P\matrixfont{P}.
\end{equation}
Unlike $\matrixfont{P}$ and $\matrixfont{W}$ which are introduced as derivation tools rather than for practical computation, $\matrixfont{P}_W$ and $\matrixfont{W}_P$ can actually be implemented in practice. Operations that are performed by \codefont{genetIC} include up-sampling the low resolution information into the high resolution (which can be notated as $\matrixfont{P}_W^+\matrixfont{W}_P$); and down-sampling the high resolution region into a zero-padded low resolution full volume ($\matrixfont{W}_P^+\matrixfont{P}_W$). 

According to the description above, $\matrixfont{P}_W^+$ can be thought of as a zero-order or ``nearest neighbor'' interpolation from low to high resolution. For some purposes we will wish to use higher-order interpolation to form smoother fields from the underlying coarse pixelization. For brevity, we do not make a heavy distinction between different orders of interpolation in this Appendix but note that \codefont{genetIC} in practice uses a tricubic interpolation scheme, similar to that outlined by \citet{lekien2005tricubic}. For consistency, \codefont{genetIC} also implements down-sampling $\matrixfont{P}_W$  using an interpolation scheme that maintains $\left.\matrixfont{P}_W \matrixfont{P}_W\hspace{-0.3em}\right.^+ = \mathbb{I}$.

\subsection{Relationship Between Variable-resolution and Uniform-resolution Fields}\label{app:variable-resolution-to-fixed}
\label{app:derivation_fast_filter}
Existing algorithms for performing modifications, including the \cite{hoffman1991constrained} algorithm resulting from the formulation in \cite{Roth:2015wha}, are framed in terms of a uniform pixelization. We therefore need to derive a robust but computationally tractable algorithm for modifying zoom initial conditions. To start, we require an explicit analytic form for the map from white noise at varying resolution to a density-contrast field sampled at high resolution across the full simulation domain. This map can be regarded as a definition from which all algorithms derive, and is given for \codefont{genetIC} by 
\begin{equation}
\nvec{\delta}  =  \matrixfont{C}^{1/2}\matrixfont{F}_H\matrixfont{W}^{+}\nvec{n}_H + m^{-1/2} \matrixfont{C}^{1/2}\matrixfont{F}_L\matrixfont{P}^{+}\nvec{n}_L\,.
\end{equation}
In words, to obtain a density-contrast field at high resolution throughout the box we would (i) place or resample the separate white noise fields into high resolution across the full volume ($\matrixfont{W}^+$ and $m^{-1/2} \matrixfont{P}^+$ respectively); (ii) apply the appropriate high-pass $\matrixfont{F}_H$ or low-pass $\matrixfont{F}_L$ filters\footnote{In actual practice, we want to retain the high-frequency modes outside the zoom region rather than cutting them off entirely, so we actually use a modified low-pass filter $\tilde{\matrixfont{F}}_L = (\mathbb{I} - \matrixfont{W}^+\matrixfont{W}) + \matrixfont{W}^+\matrixfont{W}\matrixfont{F}_L$ where $\matrixfont{F}_L$ is the original Fermi filter. This is equivalent to applying the low-pass filter only in the zoom region, and does not significantly affect how the filters operate.}; and finally (iii) convolve with  $\matrixfont{C}^{1/2}$ to obtain an appropriate covariance. \footnote{In general, the covariance of $\nvec{\delta}$ as obtained from $\nvec{n}_H$ and $\nvec{n}_L$ is not precisely $\matrixfont{C}$, because outside the zoom region, high-frequency modes contributing to $\matrixfont{C}$ cannot be represented. The exact covariance $\matrixfont{C}$ is only obtained in the limit that the zoom region occupies the entire space, $\matrixfont{W}^+ \matrixfont{W}=\mathbb{I}$ and that $\matrixfont{F}_L$ band limits the signal in the coarse pixelization, $\matrixfont{P}^+ \matrixfont{P} \matrixfont{F}_L = \matrixfont{F}_L$.} 

For compact notation, one can concatenate the two noise vectors $\nvec{n}_L$ and $\nvec{n}_H$ into a single vector $\nvec{n}_Z$ (where $Z$ stands for ``zoom''), and write 
\begin{equation}
\nvec{\delta} = \matrixfont{T} \nvec{n}_Z\,, \quad \textrm{ where } \matrixfont{T} = \left( \begin{matrix}
	\matrixfont{C}^{1/2}\matrixfont{F}_H\matrixfont{W}^{+} &\quad  m^{-1/2} \matrixfont{C}^{1/2}\matrixfont{F}_L\matrixfont{P}^{+}\label{eq:coordTrans}
\end{matrix}
\right)\,.
\end{equation}
Thus, the transformation matrix $\matrixfont{T}$ is a $N \times (N_P + N_W)$ matrix. We never explicitly calculate $\nvec{\delta}$ or indeed $\matrixfont{T}$, since these are prohibitively large in realistic simulations; its form is required only in the derivation of algorithms.

Given any matrix transformation $\matrixfont{M}$ that can be applied to a uniform resolution volume $\nvec{\delta}$, we would ideally wish to find an equivalent matrix $\matrixfont{M}_Z$ that applies to the variable-resolution white noise, such that 
\begin{equation}
	\matrixfont{MT}\nvec{n}_Z = \matrixfont{TM}_Z \nvec{n}_Z\label{eq:define-matrix-transform},
\end{equation}
for any vector $\nvec{n}_Z$. In general, there is not an exact solution to this problem because $\matrixfont{T}$ is non-invertible, meaning that $\matrixfont{M}_Z$ is overdetermined. However, a uniquely well-motivated approximation is obtained by taking
\begin{equation}
\matrixfont{M}_Z = \matrixfont{T}^{+}\matrixfont{M}\matrixfont{T}\,\label{eq:matrix-transform},
\end{equation}
where $\matrixfont{T}^+$ is the pseudo-inverse of $\matrixfont{T}$. The motivation for this expression can be understood at three levels of detail: 
\begin{enumerate}
	\item It is the most obvious generalization of the standard matrix transformation $\matrixfont{M}_Z = \matrixfont{T}^{-1} \matrixfont{M} \matrixfont{T}$ to the case where $\matrixfont{T}$ is non-invertible.
	\item  In the case that $\matrixfont{T}^\dagger \matrixfont{T}$ is invertible, expression~\eqref{eq:matrix-transform} can be derived exactly from~\eqref{eq:define-matrix-transform}, using the pseudo-inverse identity $\matrixfont{T}^+ \equiv (\matrixfont{T^\dagger T})^{-1} \matrixfont{T}^\dagger$. 
	\item In the case of interest, where neither $\matrixfont{T}$ nor $\matrixfont{T}^\dagger \matrixfont{T}$ are expected to be invertible, the expression can instead be derived using a maximum likelihood principle. This is shown in Appendix~\ref{app:ML-pseudo-inverse}.
\end{enumerate}

In order to compute $\matrixfont{M}_Z$ from $\matrixfont{M}$ and hence formulate a practical implementation of modifications in zoom simulations, we will need an explicit expression for $\matrixfont{T}^+$.  It is easiest to understand its form by bearing in mind that, practically speaking, $\matrixfont{T}^+$ maps from the uniform resolution $\vec{\delta}$ back onto varying-resolution white-noise fields $\vec{n}_Z$. With this in mind, one can guess at an approximate solution,
\begin{equation}
\nvec{n}_Z = \matrixfont{T}^+ \nvec{\delta}\,, \quad \textrm{where } \matrixfont{T}^{+} = \left(\begin{matrix}\matrixfont{W}\matrixfont{F}_H\matrixfont{C^{-1/2}}\\m^{1/2}\matrixfont{P}\matrixfont{F}_L\matrixfont{C^{-1/2}}\end{matrix}\right).\label{eq:Tplus-guess}
\end{equation}
The $\matrixfont{C}^{-1/2}$ operators deconvolve and thus restore the white noise property of the original noise fields. Then, for the high resolution grid, a high-pass filter is applied ($\matrixfont{F}_H$) and the appropriate region is extracted ($\matrixfont{W}$).  For the low resolution grid, the low-pass filter is applied ($\matrixfont{F}_L$) and the entire domain downsampled ($m^{1/2} \matrixfont{P}$, with the $m^{1/2}$ factor serving to preserve the unit variance of the white noise). Further discussion and motivation of this approximate pseudo-inverse is given in Appendix~\ref{app:properties-approximate-pseudo-inverse}.

\subsection{Practical Implementation: Generating Zoom Initial Conditions}
\label{sec:simplify}
We next describe how the defining relation~\eqref{eq:coordTrans} relates to the practical computation of $\vdelta_H$ at high resolution in the zoom window and $\vdelta_L$ at low resolution across the full box.  
For consistency, we should expect that $\nvec{\delta}_{H} = \matrixfont{W}\nvec{\delta}$, i.e. that the high-resolution portion of our final overdensity field is given by extracting the relevant part of $\vdelta$. However, as previously discussed, a direct computation of $\vdelta$ is prohibitive so we are forced to make some approximations. First, let $\matrixfont{F}_{WH}$ be the high-pass filter defined on the windowed region only. The filter is chosen to remove modes near the fundamental mode of the zoom window, as discussed in Section~\ref{sec:fast_filter}. Consequently we may assume that
\begin{equation}
\matrixfont{W}\matrixfont{F}_{H} \approx \matrixfont{F}_{WH}\matrixfont{W}\,,\label{eq:Wassumption}
\end{equation}
i.e. high-pass filtering and extraction of the zoom window approximately commute. 
 Additionally, we assume
\begin{equation}
\matrixfont{F}_L\matrixfont{P}^{+} \approx \matrixfont{P}^{+}\matrixfont{F}_{PL}\label{eq:Passumption},
\end{equation}
where $\matrixfont{F}_{PL}$, is the low-pass filter acting on the pixelized grid. This approximation corresponds to the assumption that the low-pass filter band-limits signals sufficiently far below the Nyquist mode of the pixelized grid, again discussed in Section~\ref{sec:fast_filter}. With these two assumptions, the field in the high-resolution region, $\nvec{\delta}_H$, can be approximated as 
\begin{equation}
\nvec{\delta}_H = \matrixfont{WT}\nvec{n}_Z \approx  \matrixfont{F}_{WH}\matrixfont{W}\matrixfont{C}^{1/2}\matrixfont{W}^{+}\nvec{n}_H+ m^{-1/2} \matrixfont{W}\matrixfont{C}^{1/2}\matrixfont{P}^{+}\matrixfont{F}_{PL}\nvec{n}_L.
\end{equation}
This expression is not suitable for practical computations, however, because $\matrixfont{F}_{WH}\matrixfont{W}\matrixfont{C}^{1/2}\matrixfont{W}^{+}$ still involves a high-resolution, full-volume convolution. We approximate this operation by applying the covariance matrix evaluated at high resolution but only in the zoom region, 
\begin{equation} 
	\matrixfont{F}_{WH}\matrixfont{W}\matrixfont{C}^{1/2}\matrixfont{W}^{+} \simeq \matrixfont{F}_{WH}\matrixfont{C}_H^{1/2}.
\end{equation} 
Because of the filter $\matrixfont{F}_{WH}$, only high-$k$ modes (far above the fundamental frequency of the zoom region) are retained after the convolution; consequently, the approximation should be excellent. 

Similarly, the operation $\matrixfont{W}\matrixfont{C}^{1/2}\matrixfont{P}^{+}$ must be replaced: naively, it would involve resampling the volume to high resolution ($\matrixfont{P}^+$), convolving, then extracting only the high resolution part ($\matrixfont{W}$). Instead, we use the approximation  
\begin{equation}
\matrixfont{W}\matrixfont{C}^{1/2}\matrixfont{P}^{+} \approx \matrixfont{P}_W^+\matrixfont{W}_P\matrixfont{C}_{L}^{1/2}.
\end{equation}
which describes convolving at low resolution ($\matrixfont{C}_{L}^{1/2}$), extracting the zoom region ($\matrixfont{W}_P$), and resampling only that region to the high resolution ($\matrixfont{P}_W^+$).  

As all of these operations are now tractable (i.e. either they are sampled at low resolution, or only encompass the zoom region), they can be efficiently implemented.  We thus arrive at the practical estimator used by \codefont{genetIC},
\begin{align}
\nvec{\delta}_H =& \matrixfont{F}_{WH}\matrixfont{C}_{H}^{1/2}\nvec{n}_H+ m^{-1/2} \matrixfont{P}_W^+\matrixfont{W}_P\matrixfont{F}_{PL}\matrixfont{C}_L^{1/2}\nvec{n}_L,\label{eq:deltaW}
\end{align}
which was illustrated in Figure~\ref{fig:zoomset up}. The coarsely pixelated field for the rest of the simulation, $\nvec{\delta}_L$, is obtained by ignoring the irrelevant high-$k$ modes in $\nvec{n}_H$:
\begin{align}
	\nvec{\delta}_L =& \matrixfont{C}_L^{1/2}\nvec{n}_L\,,\label{eq:deltaP}
\end{align}
where $\matrixfont{C}_L$ is the covariance matrix evaluated at low resolution. 

At the time that the overdensity field is computed we also calculate the velocity field and hence Zel'dovich displacements, which are required for generating particle output. These are obtained using precisely the algorithm above, but using the appropriate covariance matrix, as described in Section~\ref{sec:othermods}. 

\subsection{Practical Implementation: Performing Modifications}
\label{app:constraints}

We are now in a position to derive the algorithm for making modifications to zoom initial conditions. In Section~\ref{sec:linear} we described how, in cases where the uniform high-resolution overdensity $\nvec{\delta}$ is available,  modifications  are defined via a covector $\nvec{u}$ and a target value $d$. The aim is to generate $\nvec{\delta}'$ which is statistically as close as possible to $\nvec{\delta}$, but which satisfies
\begin{equation}
\nvec{u}\cdot \nvec{\delta}' = d.
\end{equation}
For these constant resolution vectors the appropriate linear transformation is given by the Hoffman-Ribak algorithm,
\begin{equation}
\nvec{\delta}' = \nvec{\delta} + \frac{(d - \nvec{u}\cdot\nvec{\delta})\matrixfont{C}\nvec{u}}{\nvec{u}\cdot \matrixfont{C}\nvec{u}}.\label{eq:HRtransformation}
\end{equation}
To produce an implementation which works at variable resolution, we first need to find a covector $\nvec{u}_Z$ satisfying
\begin{equation}
\nvec{u}_Z\cdot\nvec{n}_Z = \nvec{u}\cdot \nvec{\delta},
\end{equation}
for the variable-resolution white noise $\nvec{n}_Z$ which generates the overdensity field $\nvec{\delta}$. A general solution can be found by substituting equation~\eqref{eq:coordTrans}, yielding
\begin{equation}
\nvec{u}_Z = \matrixfont{T}^{\dagger}\nvec{u}.\label{eq:uZ-from-u}
\end{equation}
This will however, be difficult to compute since it starts from the full high-resolution vector $\nvec{u}$ which by assumption cannot be stored. To simplify, we choose to consider only modifications where the objective is specified within the high resolution region\footnote{The resulting modifications will still affect the low resolution region, as we continue to include the covariance across the entire simulation domain. This is clear, for example, in Figure~\ref{fig:velMod}.}.  For scientific applications, this is naturally the case --- the highest resolution is, by construction, centered around the objects of interest. 

Such covectors will satisfy $\matrixfont{W}^{+}\matrixfont{W}\nvec{u} = \nvec{u}$ since they are zero outside the high-resolution region. This means that we can replace \eqref{eq:uZ-from-u} by a feasible computation,
\begin{equation}
\nvec{u}_Z = (\matrixfont{W}\matrixfont{T})^{\dagger}(\matrixfont{W}\nvec{u}),\label{eq:uz}
\end{equation}
where we start from $\matrixfont{W}\nvec{u}$ which is the $\nvec{u}$ covector calculated only within the high resolution region.  
We previously described the approximations used for calculating $\vdelta_H = \matrixfont{WT} \nvec{n}_Z$, and by using the same approximations we can write $(\matrixfont{WT})^\dagger$ in a tractable form: 
\begin{equation}
(\matrixfont{W}\matrixfont{T})^{\dagger} \approx \left(\begin{matrix}\matrixfont{F}_H\matrixfont{C}_W^{1/2}\\m^{1/2}\matrixfont{F}_{PL}\matrixfont{C}_P^{1/2}\matrixfont{W}_P^+\matrixfont{P}_{W}^{+\dagger}\end{matrix}\right).\label{eq:WT-dagger-approx}
\end{equation}
We must also transform the covariance matrix $\matrixfont{C}$ appearing in equation (\ref{eq:HRtransformation}) to obtain $\matrixfont{C}_Z$ according to the approximate solution~\eqref{eq:coordTrans}, where $\matrixfont{T}^+$ is specified by Equation~\eqref{eq:Tplus-guess}. This leads to the result
\begin{equation}
\matrixfont{C}_Z = \left(\begin{matrix}\matrixfont{W}\matrixfont{F}_H^2\matrixfont{W}^{+}&m^{-1/2}\matrixfont{W}\matrixfont{F}_H\matrixfont{F}_L\matrixfont{P}^{+}\\m^{1/2}\matrixfont{P}\matrixfont{F}_L\matrixfont{F}_H\matrixfont{W}^{+}&\matrixfont{P}\matrixfont{F}_L^2\matrixfont{P}^{+}\end{matrix}\right).
\end{equation}
Once again, this cannot be implemented directly because it involves operations defined at high-resolution over the whole simulation. Applying the same approximations used in obtaining~\eqref{eq:WT-dagger-approx}, so that all operations are either performed in the zoom window or at low resolution, we find the appropriate covariance matrix to be
\begin{equation}
\matrixfont{C}_Z \approx \left(\begin{matrix}\matrixfont{F}_{WH}^2&m^{-1/2}\matrixfont{F}_{WH}\matrixfont{F}_{WL}\matrixfont{W}_P\matrixfont{P}_W^+\\m^{1/2}\matrixfont{F}_{PL}\matrixfont{F}_{PH}\matrixfont{P}_{W}\matrixfont{W}_P^+&\matrixfont{F}_{PL}^2\end{matrix}\right).
\end{equation}
For completeness, we now write the updated Hoffman-Ribak transformation as
\begin{equation}
\nvec{n}_Z' = \nvec{n}_Z  + \frac{(d - \nvec{u}_Z\cdot\nvec{n}_Z)\matrixfont{C}_Z\nvec{u}_Z}{\nvec{u}_Z\matrixfont{C}_Z\nvec{u}_Z}.\label{eq:zoom-hoffman-ribak}
\end{equation}

When applying multiple modifications, as explained by \cite{Roth:2015wha}, we apply Gram-Schmidt orthogonalization and then are able to treat each as independent. The orthogonalization process makes use of the covariance $\matrixfont{C}$, which must be replaced by $\matrixfont{C}_Z$ in the case of zoom simulations. Constraints on the potential or velocity fields can be implemented by replacing all instances of the covariance matrix $\matrixfont{C}$ with the appropriately re-weighted matrix and re-deriving $\matrixfont{C}_Z$, as described in Section~\ref{sec:othermods}. Because the case of quadratic modifications \citep{2018MNRAS.474...45R} is linearized and turned into an iterative set of linear modifications, they too can be handled naturally using the updated transformation law~\eqref{eq:zoom-hoffman-ribak}.

\section{Technical Details of Approximations}\label{app:even-more-technical}

\subsection{A Maximum Likelihood Derivation of the Pseudo-inverse}\label{app:ML-pseudo-inverse}

In Appendix~\ref{app:derivation_fast_filter}, we discussed how modifying the initial conditions for zoom simulations requires us to find matrices $\matrixfont{M}_Z$ which satisfy $\matrixfont{MT}\nvec{n}_Z = \matrixfont{TM}_Z \nvec{n}_Z$. In general, there is no exact solution to the equation $\matrixfont{MT} = \matrixfont{TM}_Z$ because $\matrixfont{T}$ is non-invertible and  because careful analysis shows that $\matrixfont{T^\dagger T}$ is also non-invertible. 

Note that this is a generalized version of the problem considered by \cite{Penrose_1956}, where a solution is sought for the vector $\vec{v}$ in the equation $\matrixfont{T} \vec{v} = \vec{u}$; in that case, $\vec{v} \simeq \matrixfont{T}^+ \vec{u}$ is a uniquely motivated approximate solution. We can similarly argue that there is a uniquely motivated choice for obtaining an approximate solution for $\matrixfont{M}_Z$. We start by considering the residual error for any candidate $\matrixfont{M}_Z$. Since we are describing modifications to $\vdelta$, which is Gaussian-distributed with covariance $\matrixfont{C}$, the error can most naturally be quantified in terms of a $\chi^2$ measure: 
\begin{equation}
\chi^2 = \frac{1}{2}\,(\matrixfont{MT}\nvec{n}_Z - \matrixfont{TM}_Z \nvec{n}_Z)^\dagger \matrixfont{C}^{-1} (\matrixfont{MT}\nvec{n}_Z - \matrixfont{TM}_Z \nvec{n}_Z)\,.
\end{equation}
We now take the expectation value across the ensemble of $\nvec{n}_Z$ (which has unit variance), yielding 
\begin{equation}
\langle \chi^2 \rangle = \frac{1}{2} \mathrm{Tr}\, \left\{ \matrixfont{C}^{-1} (\matrixfont{MT} - \matrixfont{TM}_Z) (\matrixfont{MT} - \matrixfont{TM}_Z)^{\dagger}\right\}\textrm{.}
\end{equation}
We wish to minimize this expected error, or in other words maximize the likelihood. We guard against poorly conditioned matrices $\matrixfont{M}_Z$ by introducing a penalty term $\mathrm{Tr}\,\{\matrixfont{M}_Z \matrixfont{M}_Z ^{\dagger}\}$ with a weighting $\alpha$ \citep[this procedure is sometimes known as Tikhonov regularization;][]{tikhonov2013numerical}. The problem is then to find the elements of $\matrixfont{M}_Z$ which minimize 
\begin{equation}
	\mathcal{F} \equiv \mathrm{Tr}\, \left\{\matrixfont{C}^{-1} (\matrixfont{MT} - \matrixfont{TM}_Z) (\matrixfont{MT} - \matrixfont{TM}_Z)^{\dagger}\right\} + \alpha \mathrm{Tr}\,\{\matrixfont{M}_Z \matrixfont{M}_Z ^{\dagger}\} \textrm{.}
\end{equation}
This is achieved by solving $\partial \mathcal{F}/\partial \matrixfont{M}_Z = 0$, leading to the expression 
\begin{equation}
	\matrixfont{M}_Z = \left( \matrixfont{T^\dagger C}^{-1} \matrixfont{T} + \alpha \mathbb{I} \right)^{-1} \matrixfont{T^\dagger C}^{-1} \matrixfont{MT}\textrm{.}
\end{equation}
We now make a temporary substitution, $\tilde{\matrixfont{T}} = \matrixfont{C}^{-1/2}\matrixfont{T}$ to obtain the simplified expression
\begin{equation}
	\matrixfont{M}_Z = \left( \tilde{\matrixfont{T}}\matrixfont{^\dagger} \tilde{\matrixfont{T}} + \alpha \mathbb{I} \right)^{-1} \tilde{\matrixfont{T}}\matrixfont{^\dagger C}^{-1/2} \matrixfont{MT}\textrm{.}
\end{equation}
Finally, we take the result in the limit that our penalty term is always subdominant, i.e. $\alpha \to 0$. Using the formal definition of the Moore-Penrose pseudo-inverse \citep{penrose_1955} for $\tilde{\matrixfont{T}}$, namely 
\begin{equation}
	\tilde{\matrixfont{T}}^+ \equiv \lim_{\alpha \to 0+} \left( \tilde{\matrixfont{T}}\matrixfont{^\dagger }\tilde{\matrixfont{T}} + \alpha \mathbb{I} \right)^{-1} \tilde{\matrixfont{T}}^{\dagger} \,.
\end{equation}
one obtains the solution, 
\begin{equation}
	\matrixfont{M}_Z = \tilde{\matrixfont{T}}^+ \matrixfont{C}^{-1/2} \matrixfont{MT} = \matrixfont{T^+ MT}\,,
\end{equation}
where we have used $\tilde{\matrixfont{T}}^+ = \matrixfont{T}^+ \matrixfont{C}^{1/2}$. This concludes the demonstration that equation~\eqref{eq:matrix-transform} is the minimum-error, i.e. maximum likelihood solution. A similar argument also shows that the maximum likelihood reconstruction of $\nvec{n}_Z$ starting from a given $\nvec{\delta}$ is given by $\nvec{n}_Z = \matrixfont{T}^+ \nvec{\delta}$.

\subsection{Properties of our Approximate Pseudo-inverse}\label{app:properties-approximate-pseudo-inverse}

In Appendix~\ref{app:derivation_fast_filter}, we commented that it was not possible to find an exact expression for $\matrixfont{T}^+$.  We instead motivated the approximate solution~\eqref{eq:Tplus-guess}. We now provide further information about this approximation and its properties.

First we note that an exact pseudo-inverse satisfies the relations
\begin{equation} 
	\matrixfont{T}^+ \matrixfont{T} \matrixfont{T}^+ = \matrixfont{T}^+ \quad \textrm{and} \quad \matrixfont{T} \matrixfont{T}^+ \matrixfont{T} = \matrixfont{T}\,,
\end{equation}
and hence
\begin{equation} 
	(\matrixfont{T} \matrixfont{T}^+)^2 = \matrixfont{T}\matrixfont{T}^+\quad\textrm{and}\quad(\matrixfont{T}^+ \matrixfont{T})^2 = \matrixfont{T}^+\matrixfont{T}.\label{eq:pseudo-inverse-projectors}
\end{equation}
This result shows $\matrixfont{T} \matrixfont{T}^+$ and $\matrixfont{T}^+ \matrixfont{T}$ should both act as projection matrices. The projections can be interpreted as follows: $\matrixfont{T}\matrixfont{T}^+ \vdelta$ takes a general overdensity field $\vdelta$ and projects out those modes that cannot be represented in our compressed scheme (i.e. high frequency modes lying outside the zoom region).  Conversely, $\matrixfont{T}^+\matrixfont{T} \nvec{n}_Z$ takes any realization of the white noise $\nvec{n}_Z$ and projects out those modes which are not used in constructing the physical overdensity field $\vdelta$. 

One way to test our approximation for $\matrixfont{T}^+$ is to check how closely the projection properties~\eqref{eq:pseudo-inverse-projectors} are adhered to. 
To start, we show that both requirements are exact in an artificial limit where the zoom region covers the entire volume; in this case, the \codefont{genetIC} algorithm continues to split information into high-frequency and low-frequency components, but without any compression. In other words, no pixel downsampling takes place and the high frequency components are retained across the entire simulation.  We thus take $\matrixfont{W}=\matrixfont{P}=\mathbb{I}$ and $m=1$, finding that 
\begin{equation}
	\matrixfont{T^+T} = \left(\begin{matrix}
		\matrixfont{F}_H^2 & \matrixfont{F}_H \matrixfont{F}_L \\
		\matrixfont{F}_L \matrixfont{F}_H & \matrixfont{F}_L^2 \\
	\end{matrix}\right) \quad \textrm{and} \quad \matrixfont{T T^+} = \mathbb{I}\,.
\end{equation}
The latter follows from the defining relation between the low-pass and high-pass filters, $\matrixfont{F}_L^2 + \matrixfont{F}_H^2 = \mathbb{I}$, previously stated in Equation~\eqref{eq:filter-completeness}. From these results, the requirements~\eqref{eq:pseudo-inverse-projectors} follow immediately, confirming that our claimed pseudo-inverse is exactly correct.

Let us now consider the more realistic case, where $\matrixfont{W}$ and $\matrixfont{P}$ are restored, i.e. high frequency information is retained only within the zoom region. We now find our approximate pseudo-inverse yields
\begin{equation}
\matrixfont{T}\matrixfont{T}^{+} = \matrixfont{C}^{1/2}\left(\matrixfont{F}_H\matrixfont{W}^{+}\matrixfont{W}\matrixfont{F}_H + \matrixfont{F}_L\matrixfont{P}^{+}\matrixfont{P}\matrixfont{F}_L\right)\matrixfont{C}^{-1/2}.
\end{equation}
By using the filter relation~\eqref{eq:filter-completeness} and factorizing the resulting terms, we find  
\begin{equation}
(\matrixfont{T}\matrixfont{T}^{+})^2 = \matrixfont{T}\matrixfont{T}^{+} + \matrixfont{C}^{1/2}(\matrixfont{F}_H\matrixfont{W}^{+}\matrixfont{W}\matrixfont{F}_L - \matrixfont{F}_L\matrixfont{P}^{+}\matrixfont{P}\matrixfont{F}_H)(\matrixfont{F}_L\matrixfont{W}^{+}\matrixfont{W}\matrixfont{F}_H - \matrixfont{F}_H\matrixfont{P}^{+}\matrixfont{P}\matrixfont{F}_L)\matrixfont{C}^{-1/2}.
\end{equation}
To obtain an exact projection, it is therefore sufficient that 
\begin{equation}
\matrixfont{F}_L\matrixfont{W}^{+}\matrixfont{W}\matrixfont{F}_H - \matrixfont{F}_H\matrixfont{P}^{+}\matrixfont{P}\matrixfont{F}_L = 0\label{eq:TplusApprox}.
\end{equation}
Although this equation will not hold in general, there are a number of reasons why it holds to sufficient accuracy for our purposes. First, it involves products of the low-pass and high-pass filters, which when composed produce a narrow band-pass effect. Thus only a small fraction of modes will be affected by the approximation. 

More technically, we can study the properties of overdensity fields  $\nvec{\delta}$ for which the operator does obey the desired relation $(\matrixfont{T}\matrixfont{T}^{+})^2 \vdelta = (\matrixfont{T}\matrixfont{T}^{+}) \vdelta$.  By expanding, one finds that $\nvec{\delta}$ must satisfy
\begin{align}
	\matrixfont{P}^{+}\matrixfont{P}\matrixfont{F}_L\matrixfont{C}^{-1/2}\nvec{\delta} =& \matrixfont{F}_L\matrixfont{C}^{-1/2}\nvec{\delta}\label{eq:Pcond}, \\
\matrixfont{W}^{+}\matrixfont{W}\matrixfont{F}_H\matrixfont{C}^{-1/2}\nvec{\delta} =& \matrixfont{F}_H\matrixfont{C}^{-1/2}\nvec{\delta}\label{eq:Wcond}.
\end{align}
The first of these relations, \eqref{eq:Pcond}, is always approximately satisfied because, by assumption, $\matrixfont{P}^{+}\matrixfont{P}\matrixfont{F}_L \simeq \matrixfont{F}_L$; that is, the entire point of the low-pass filter is to band-limit such that the coarse pixelization becomes irrelevant.  The second relation, equation (\ref{eq:Wcond}), states that we must start with overdensity fields that have high frequency information only within the zoom region. These are precisely the properties we would expect of fields that can be represented accurately in zoom initial conditions.

The operator $\matrixfont{T}^{+}\matrixfont{T}$, on the other hand, acts on $\nvec{n}_Z$ and takes the form
\begin{equation}
\matrixfont{T}^{+}\matrixfont{T} = \left(\begin{matrix}\matrixfont{W}\matrixfont{F}_H^2\matrixfont{W}^{+}&m^{-1/2}\matrixfont{W}\matrixfont{F}_H\matrixfont{F}_L\matrixfont{P}^{+}\\m^{1/2}\matrixfont{P}\matrixfont{F}_L\matrixfont{F}_H\matrixfont{W}^{+}&\matrixfont{P}\matrixfont{F}_L^2\matrixfont{P}^{+}\end{matrix}\right).
\end{equation}
We find the conditions for this to be a projection operator are the same as for $\matrixfont{T} \matrixfont{T}^+$, i.e. they are given by equation~(\ref{eq:TplusApprox}).  Thus for fields which are of interest to \codefont{genetIC}, we expect the approximation to be excellent, as borne out by the tests in Section~\ref{sec:examples}.

\bibliographystyle{aasjournal}
\bibliography{codeRelease}

\end{document}